\documentclass[preprint,12pt]{elsarticle}

\usepackage{amssymb}

\def\linkDir{links/}
\def\bibFile{\linkDir/ref/refEx.bib}
\biboptions{sort&compress}

\usepackage{amsmath,amsfonts,amsthm,bm,amssymb}
\usepackage{mathtools}
\usepackage{wasysym}
\usepackage{upgreek}
\usepackage{BOONDOX-cal}

\usepackage{multirow}
\usepackage{booktabs}
\usepackage{longtable}

\usepackage{graphicx}
\usepackage{overpic}
\usepackage{color}
\usepackage[utf8]{inputenc}

\usepackage{caption}

\usepackage{paralist}

\usepackage{framed}

\usepackage{tikz}
\usetikzlibrary{positioning}
\usetikzlibrary{shapes,arrows,arrows,positioning,fit}

\usepackage{wasysym}
\usepackage{textcomp}
\usepackage{pdflscape}

\usepackage{hyperref} 
\hypersetup{colorlinks,
            linkcolor=blue,
            citecolor=blue,
            urlcolor=blue,
            pdfauthor=DeAn Wei}

\usepackage{cleveref}
\crefname{equation}{Eq.}{Eqs.}
\crefname{figure}{Fig.}{Figs.}

\def\bi#1{\textbf{\emph{#1}}}

\usepackage{soul}

\usepackage[section]{placeins}
\usepackage{float}
\newcommand \MZ [1] {\bgroup\noindent\textcolor{blue}{[\textbf{MZ}: #1]}\egroup\ignorespacesafterend}

\journal{Journal of XXX}

\begin{document}

\begin{frontmatter}



\title{Discrete Dislocation Dynamics Modeling of Nanotwinned Materials I: Orientation Effects in a Multilayer Twinned Structure of Copper}

\author[swjtu]{DeAn Wei}
\author[fau]{Michael Zaiser}
\author[lxs]{Jing Tang}

\author[swjtu]{Xu Zhang\corref{cor1}}
\ead{xzhang@swjtu.edu.cn}

\affiliation[swjtu]{organization={Applied Mechanics and Structure Safety Key Laboratory of Sichuan Province, School of Mechanics and Aeronautics}, 
            city={Chengdu},
            postcode={610031}, 
            state={Sichuan},
            country={China}}

\affiliation[lxs]{organization={Institute of Mechanics, Chinese Academy of Sciences}, 
            city={Beijing},
            postcode={100190}, 
            country={China}}

\affiliation[fau]{organization={Department of Materials Science and Engineering, Institute of Materials Simulation WW8, Friedrich-Alexander University of Erlangen-Nuremberg}, 
            addressline={Dr.-Mack-Str.77}, 
            city={Fuerth},
            postcode={90762}, 
            country={Germany}}

\cortext[cor1]{Corresponding author}

\begin{abstract}
The impact of twin boundaries (TBs) on the microstructure evolution and plastic deformation mechanisms of face-centered cubic (FCC) metals has been extensively studied since the discovery that nanotwinned materials exhibit a favorable combination of high strength and ductility. In this work, a dislocation-twin boundary interaction model for copper is incorporated into a three-dimensional discrete dislocation dynamics (DDD) framework. This approach is applied to systematically investigate the orientation effects on the deformation of nanotwinned copper, utilizing a multilayer twinned structure (MTS) with a twin thickness of 160 nm.

The simulation results show that the stress-strain response of MTSs under uniaxial loading depends significant on the orientation of the loading axis. Dislocations inclined to TBs are confined to slip in single- or multi-layer twin lamellae; when the loading axis is oriented perpendicular or parallel to TBs, such  whereas when loading axis inclined to TBs, the dislocations with glide plane parallel to TBs are easily activated (mainly twinning dislocations) and the TBs do not hinder the dislocations and behave in soft modes.
If the hard mode dominates the deformation mechanism, microstructures with single-layer confined slip lead to significant hardening behavior, while microstructures with multilayer confined slip maintain stable plastic flow and do not lead to hardening.
Finally, through the introduction of critical resolved shear stresses (CRSSs) specific to various deformation modes and the adaptation of Schmid's law, we have effectively projected the additional anisotropic characteristics induced by TBs in MTSs.

\end{abstract}



\end{frontmatter}


\section{Introduction}
During the past 20 years, nanotwinned materials have attracted significant interest due to their favorable combination of strength and ductility \cite{Li-Dislocation-nucleation-Nature-2010,Lu-Strengthening-Materials-Science-2009,Lu-Stabilizing-nanostructures-Nature-2016,Pan-History-independent-cyclic-Nature-2017,Wei-Evadingstrengthductility-NC-2014,Cheng-Extra-strengthening-Science-2018}. The good mechanical properties can be understood in terms of the interaction between dislocations and twin boundaries: TBs not only act as dislocation obstacles similar to conventional grain boundaries (GBs), resulting in a GB strengthening effect \cite{You-Tensile-behavior-Acta-2011,Zhang-Nanoscale-twinning-induced-strengthening-Applied-2004,Duan-Ultrastrongnanotwinnedpure-SA-2021}, but can also accommodate dislocation glide and act as dislocation reaction sites, thus ensuring sufficient work hardening to prevent failure by strain localization \cite{Chen-Size-dependence-Scripta-2011,Lu-Nano-sized-twins-Acta-2005,Wang-Sliding-of-Nature-2017}.

Unlike traditional FCC polycrystals, nanotwinned materials have anisotropic intragranular microstructures characterized by a high density of twins that directly break the equivalence of the 12 slip systems of the FCC structure \cite{Lu-Plastic-deformation-Acta-2014}. This causes significant anisotropy in dislocation motion, making the process of plastic deformation highly dependent on grain orientation \cite{You-Tensile-behavior-Acta-2011,Li-Dislocation-nucleation-Nature-2010,Lu-Size-dependence-Scripta-2009}. Based on the different types of interactions between dislocations and TBs, the slip systems in twinned structures can be categorized into the following modes \cite{Lu-Stabilizing-nanostructures-Nature-2016,Lu-Plastic-deformation-Acta-2014}: 
\begin{enumerate}[(i)]
    \item Hard mode I. In this mode, the dislocations move on glide planes that are inclined to the TBs, and the line of intersection between the twin boundary and the glide plane is at an angle of $60^{\circ}$ to the dislocation Burgers vector. In situ experiments  \cite{Wang-Atomic-scale-in-Applied-2009} have demonstrated that the $60^{\circ}$ dislocation can be transmitted through the TB by changing its Burgers vector, leaving a twinning dislocation (TD) in the TB. Molecular dynamics (MD) simulations \cite{Jin-Interactions-between-Acta-2008} predict reaction stresses as high as 1 GPa, which implies that the TBs in this mode form strong obstacles to dislocation motion.
    \item Hard mode II. In this mode, the dislocations move on glide planes that are inclined to the TBs, and the intersection lines of the TBs and the glide planes are parallel to the Burgers vector such that the dislocation is of screw type when impinging on a twin boundary. According to MD simulations, impinging screw dislocations can either be completely absorbed by the TBs and decompose into pairs of TD \cite{Zhu-Interfacial-plasticity-Proceedings-2007,Chassagne-Atomic-scale-simulation-Acta-2011,Dupraz-Large-scale-Acta-2019,Asaro-Are-rate-Scripta-2008}, forming in MTS confined hairpin-like patterns \cite{Zhu-Strengthening-mechanisms-International-2015,Zhou-A-jogged-Nano-2014,Zhou-A-plastic-JAM-2015}, or they can be transmitted to the mirror glide plane \cite{Zhu-Interfacial-plasticity-Proceedings-2007,Chassagne-Atomic-scale-simulation-Acta-2011,Dupraz-Large-scale-Acta-2019,Asaro-Are-rate-Scripta-2008,Jin-The-interaction-Scripta-2006}, leading in MTS to necklace-like patterns \cite{Zhu-Strengthening-mechanisms-International-2015,Zhou-A-jogged-Nano-2014,Zhou-A-plastic-JAM-2015}.
    \item Soft mode. Both the glide plane and the Burgers vector of soft mode dislocations are parallel to the TBs and these dislocations may glide freely even if only comparatively low resolved shear stresses are acting in the respective slip systems. At the same time it may be noted that in-plane shear stresses acting on TBs may activate an additional mode of deformation: Under these circumstances, TDs are easily emitted from the intersections of TBs and GBs, which leads to a twinning mode of deformation mediated by motion of TDs \cite{Wang-In-situ-Philosophical-2007}. Both dislocations in soft modes and twinning modes of deformation, if activated, may contribute significantly to the ductility of nanotwinned materials.
\end{enumerate}

To study grain orientation effects in nanotwinned materials, You et al. \cite{You-Tensile-behavior-Acta-2011} used direct current electrodeposition to synthesize nanotwinned materials with columnar microstructure, consisting of columnar nanotwinned grains with preferential orientation. Subsequently, You et al. \cite{You-Plastic-anisotropy-Acta-2013} investigated orientation effects in columnar nanotwinned copper by conducting tensile experiments and MD simulations. The results showed that under loading perpendicular to the TBs, hard mode I dislocations impinge on TBs and are unable to transmit, resulting in the highest yield stress; when loads are applied parallel to TBs, hard mode II dislocations are confined to slip in twin lamellae, resulting in the second highest yield stress; when loads are applied under an angle of ($45^{\circ}$) to the TBs, deformation is mediated by soft mode TDs which nucleate at low stresses and move parallel to the TBs until they gradually accumulate in front of the grain boundaries, leading to the lowest yield stress and significant strain hardening.

When You et al. \cite{You-Plastic-anisotropy-Acta-2013} explored the anisotropy of slip modes using MD, their investigation focused on columnar grains. To achieve a generalized understanding of orientation-dependent plasticity in nanotwinned structures, it is essential to isolate the geometric loading effects from texture constraints and track the sustained collective evolution of dislocation ensembles. To this end, we employ a three-dimensional discrete dislocation dynamics (DDD) framework which we implement within a simplified twin multilayer model.

DDD is a useful tool to explore the relationship between collective dislocation evolution and mechanical behavior and its dependence on microstructure parameters. 
Fan et al. \cite{Fan-The-role-Acta-2015,Fan-Orientation-influence-Scripta-2015,Fan-Grain-size-Scripta-2016} studied the plastic deformation mechanisms of magnesium alloys by introducing TBs and GBs into DDD. 
Wei et al. \cite{Wei-TMS-2019-Nature-2017,Wei-Effects-of-Acta-2019} implemented improved  dislocation-twin interaction models within a DDD framework and probed size effects and orientation effects in twinned micropillars. Based on these previous studies, this paper formulates a geometric model for a twinned multilayer structure. The focus is on the investigation of anisotropy effects arising from dislocation-TB interactions to understand the influence of TB orientation on the strength, plasticity, and work-hardening behavior of nanotwinned metals. Anisotropies related to the orientation of grain boundaries in polycrystals are  not considered. 
The study is organized as follows: Section \ref{methods-models} describes the formulation and implementation of the nanotwinned copper model within a DDD framework, Section \ref{results-discussions} presents and discusses the  simulation results, and Section \ref{Conclusions-prospects} provides conclusions and outlook.

\section{Methods and models}\label{methods-models}
The DDD platform that we used, ParaDiS (Parallel Dislocation Simulator), was developed by Lawrence Livermore National Laboratory, USA \cite{Arsenlis-Enabling-strain-Modelling-2007,Bulatov-Dislocation-multi-junctions-Nature-2006}.
Into this platform, the present authors introduced TBs within FCC structures, considering three different types of dislocation-twin boundary reactions \cite{Wei-TMS-2019-Nature-2017,Wei-Effects-of-Acta-2019}. In the following, based on previous work, a more refined model is developed and applied to nanotwinned copper.

\subsection{Forces on dislocations and local stress criteria}
The dislocation system is represented as an assembly of nodes connected by straight segments. The location of node $i$ is denoted as $\bi{x}_i$. The force $\bi{F}_i$ acting on node $i$ is assembled from the forces on the segments connected to this node \cite{Arsenlis-Enabling-strain-Modelling-2007}, i.e. $\bi{F}_i=\sum_j\bi{F}_{ij}$, where $ij$ denotes the segment connecting nodes $i$ and $j$ and $\bi{F}_{ij}$ denotes the force on this segment. The midpoint of segment ${ij}$ is denoted as $\bi{x}_{ij}= (\bi{x}_{i}+\bi{x}_{j})/2$, the segment vector is $\bi{l}_{ij} = \bi{x}_j - \bi{x}_i$, and the unit vector in segment glide direction is $\bi{m}_{ij}$.
The force $\bi{F}_{ij}$ contains the externally applied force $\bi{F}_{ij}^{\rm ext}$, the core force $\bi{F}_{ij}^{\rm c}$, the elastic self-force $\bi{F}_{ij}^{\rm es}$ and long-range interactions between the dislocation segments, which can be expressed as \cite{Arsenlis-Enabling-strain-Modelling-2007}, 
\begin{equation}
	\bi{F}_{ij}=\bi{F}_{ij}^{\rm ext}+{F}_{ij}^{\rm c}+\bi{F}_{ij}^{\rm es}+\sum_{k=1}^{N-1}\sum_{l=k+1}^N \bi{F}_{ij}^{kl}, \quad \left(\left[k,l\right]\neq\left[i,j\right]\; {\rm or} \; \left[j,i\right]\right),
	\label{nodal-force}
\end{equation}
where $N$ denotes the number of dislocation nodes and $\bi{F}_{ij}^{kl}$ denotes the long-range elastic force on the segment $ij$ by the segment $kl$.

Atomic-scale simulations and theoretical considerations \cite{Chen-Repulsive-force-Physical-2007,Deng-Repulsive-force-Scripta-2010,Zhang-Twin-Boundaries-Scientific-reports-2016,Zhang-non-Screw-TB-repulsive-JMST-2021} indicate that a lattice dislocation experiences a repulsive force when approaching a TB: Both the crystal lattice and its twin have cubic symmetry but different orientation, therefore if expressed in slip system coordinates, the lattice and its twin have different shear moduli. This leads to an image interaction with magnitude inversely proportional to the distance between the dislocation and the TB, which must be added to the forces experienced by the dislocation (note that, in the original ParaDiS code, the material is treated as isotropic, hence the image interaction is not accounted for). We evaluate the repulsive force between a TB and an approaching dislocation as a function of its distance $\cal D$ from the TB as
\begin{equation}
	f^{\rm r}=\begin{cases}
		\tau_{t}^{\rm r}b & {\cal D} > {\cal D}_{\rm c}, \\
		\frac{a\tau_t^{\rm r}b}{a-{\cal D} - {\cal D}_{\rm c}}& {\cal D} \le {\cal D}_{\rm c}.
	\end{cases}
	\label{ch2:repulsive-force-TB}
\end{equation}
The subscript $t \in \{\text{HI}, \text{HII}\}$ denotes the character of the approaching dislocation, where HI labels hard mode I dislocations and HII labels hard mode II dislocations. The parameter $a$ controls the long-distance asymptotics of the repulsive force, and the critical distance ${\cal D}_{\rm c}$ is of the order of the dislocation core width plus the effective width of the TB. Finally, the reaction stress $\tau_{\rm t}^{\rm r}$ defines the critical stress level required for dislocation transmission across the TB. 

Since the resistance to imcoming dislocations may increase sharply when they are approaching TBs, two Gauss points are adopted to improve the calculation accuracy of the repulsive force, which is evaluated as
\begin{equation}
	\bi{F}_{ij}^{\rm r} = - \frac{1}{4}\|\bi{l}_{ij}\|\bi{m}_{ij}\left[
		\bi{f}^{\rm r}\left(\bi{x}_{ij}-\frac{\bi{l}_{ij}}{2\sqrt{3}})\right) + \bi{f}^{\rm r}\left(\bi{x}_{ij}+ \frac{\bi{l}_{ij}}{2\sqrt{3}})\right)
	\right].
	\label{ch2:Gaussian-repulsive-force}
\end{equation}
After each time step, one needs to clip the crossed segments/nodes back to the TBs \cite{Wei-Effects-of-Acta-2019} and then use a local stress criterion to determine whether they react with the TBs or are transmitted. A hard mode I dislocation reacts with a TB when the local resolved shear stress $\tau^{\rm g}$ is greater than the critical reaction stress $\tau_{\rm HI}^{\rm r}$. Considering that thermal activation may reduce the critical stress found in MD simulations performed at 0K \cite{Jin-Interactions-between-Acta-2008}, the reaction stress is set to $\tau_{\rm HI}^{\rm r} = 500~{\rm MPa}$ in the current simulations. 

The determination of the reaction stress for hard mode II dislocation transmission presents a challenge due to conflicting experimental and simulation data in the published literature. First, experimental measurements on twinned bicrystals Copper grown by the Bridgman method report a relatively low transmission stress of $\sim$17 MPa \cite{Malyar-Dislocation-slip-Acta-2018}. However, it must be noted that the Bridgman technique typically produces high-quality crystals and pristine TBs, offering minimal resistance to dislocation transmission. Conversely, MD simulations involving perfect TBs predict a significantly higher reaction stress exceeding 200 MPa \cite{Dupraz-Large-scale-Acta-2019,Jin-Interactions-between-Acta-2008}. This value likely represents a theoretical upper bound, as MD simulations operate at extremely high strain rates and often neglect thermally activated processes that facilitate transmission. Crucially, the nanotwinned Cu structures modeled in this work correspond to materials typically synthesized by magnetron sputtering. As characterized by Wang et al. \cite{Wang-Defective-twin-Nature-Mat-2013}, the resulting twin boundaries are rarely geometrically perfect; they inherently harbor a high density of growth defects, such as kink-like steps. These intrinsic interfacial defects may increase the energy barrier for transmission compared to the pristine boundaries in Bridgman-grown micropillars. Explicitly modeling such discrete interface defects in DDD is computationally prohibitive. Therefore, we adopt an effective reaction stress of 100 MPa. This value is chosen to phenomenologically capture the enhanced resistance arising from the defective nature of sputtered nanotwinned boundaries and represents a compromise between the lower bound provided by experiments on pristine bicrystals and the upper bound provided by athermal MD predictions.


On the other hand, the passing reaction between a hard mode II dislocation, which is of screw type, and a TB can be envisaged as a cross slip process governed by the Friedel-Escaig mechanism \cite{Rao-Response-surface-Acta-2020,Dupraz-Large-scale-Acta-2019,Asaro-Are-rate-Scripta-2008}, hence, it not only depends on the local resolved shear stress in the dislocation slip system but also on the Escaig stress $\tau^{\rm E}$ which acts on the edge components of the two partials of the hard mode II dislocation to either narrow (negative values) or extend (positive values) the stacking-fault width. Based on the cross-slip nucleation model proposed by Rao et al. \cite{Rao-Response-surface-Acta-2020}, the contribution of Escaig stress to the cross-slip reaction in twinned copper is about $15\%$. Following this model, we evaluate the local stress for the passing of a hard mode II dislocation across a TB using the condition $\tau^{\rm g} - 0.15\tau^{\rm E} \ge \tau_{\rm HII}^{\rm r}$.

\subsection{Dislocation mobility law}

For the purpose of establishing the relationship between the nodal velocity and the Peach-Koehler forces acting on segments of both FDs and TDs, i.e., the dislocation mobility law, molecular dynamics (MD) simulations of the motion of individual dislocations have been implemented in \ref{MD-simulation}.
We first focus on the drag force $f^{\rm d}_{ij}$ in the glide direction of the segment $_{ij}$, which is balanced by the glide component of the Peach-Koehler force, i.e., $f^{\rm d}_{ij}=\tau \|\bi{b}_{ij}\|$. In general, the drag force $f^{\rm d}_{ij}$ depends on the dislocation segment velocity $v$ in a non-linear manner, see \ref{MD-simulation}. This dependency can be written in the form
\begin{equation}
	-f^{\rm d}(v)=B\left(v\right)v+f^{\rm b}\left(v\right),
	\label{cha:fdrag-v0relation-2}
\end{equation}
where $B\left(v\right)$ denotes the viscous coefficient function (slope of the tangent line)  and $f^{\rm b}\left(v\right)$ denotes the back force function, which is the intercept of the tangent line at $v=0$. The functions $B\left(v\right)$ and $f^{\rm b}\left(v\right)$ for different types of dislocations in twinned copper are obtained from the fitting functions and fit parameters given in \ref{MD-simulation}.

The net drag force on a node can be evaluated from the drag force distribution along the adjacent segments, which in turn depends on the velocities of their two nodal endpoints. Balancing forces by equating nodal forces and drag forces at all nodes thus leads to a coupled system of nonlinear equations for all nodal forces and velocities. To decouple this system, it is convenient to make the simplifying assumption (which becomes exact in the limit of infinitesimal segment length) that velocity variations are small across a segment such that we can, for the purpose of evaluating the drag force, identify the segment velocity with the nodal velocity,  $\bi{v}_i \approx \bi{v}_{ij}$. 
This velocity can be decomposed as
\begin{equation}
	\bi{v}_i=\bi{m}_{ij}\otimes\bi{m}_{ij} \cdot \bi{v}_{i}+
	 \bi{$\xi$}_{ij}\otimes\bi{$\xi$}_{ij} \cdot \bi{v}_{i}+
	 \bi{n}_{ij}\otimes\bi{n}_{ij}\cdot \bi{v}_{i}, 
	\label{ch2:velocity-decomposion}
\end{equation}
where $\bi{n}_{ij}$ and $\bi{$\xi$}$ represent the glide plane normal and the line direction of segement $ij$, respectively. The three terms on the right side of the equation above indicate the velocity components in the glide direction, the line direction and the climb direction, respectively.

For generic angles  $\beta_{ij}$ between line direction and Burgers vector direction, the viscous coefficient $\mathcal{B}_{ij}^{\rm g}(\bi{v}_{i},\beta_{ij})$ and back force $\mathcal{F}_{ij}^{\rm b,g}(\bi{v}_{i},\beta_{ij})$ for glide motion are interpolated according to 
\begin{equation}
	\begin{aligned}
		\mathcal{B}_{ij}^{\rm g}(\bi{v}_{i},\beta_{ij}) &=
		B_{\rm s}(\bi{v}_{i}\cdot \bi{m}_{ij})\cos^2\beta_{ij} + B_{\rm e}(\bi{v}_{i}\cdot \bi{m}_{ij})\sin^2\beta_{ij}, \\
		\mathcal{F}_{ij}^{\rm b,g}(\bi{v}_{i},\beta_{ij}) &=
		f_{\rm s}^{\rm b}(\bi{v}_{i}\cdot \bi{m}_{ij})\cos^2\beta_{ij} + f_{\rm e}^{\rm b}(\bi{v}_{i}\cdot \bi{m}_{ij})\sin^2\beta_{ij},
	\end{aligned}
	\label{ch2:mixed-drag-coeffient-tensor}
\end{equation}
where the subscripts $\rm s$ and $\rm e$ denote the respective values for screw and edge dislocations. We assume that similar relations also apply to nodal motion that involves a component parallel to the dislocation line, hence the respective viscous coefficients $\mathcal{B}_{ij}^{\rm l}(\bi{v}_{i},\beta_{ij})$ and back force $\mathcal{F}_{ij}^{\rm b,l}(\bi{v}_{i},\beta_{ij})$ are evaluated according to
\begin{equation}
	\begin{aligned}
		\mathcal{B}_{ij}^{\rm l}(\bi{v}_{i},\beta_{ij})&=
			B_{\rm s}(\bi{v}_{i}\cdot \bi{$\xi$}_{ij})\cos^2\beta_{ij} + B_{\rm e}(\bi{v}_{i}\cdot\bi{$\xi$}_{ij})\sin^2\beta_{ij}\\
		\mathcal{F}_{ij}^{\rm b,l}(\bi{v}_{i},\beta_{ij})&=
			f_{\rm s}^{\rm b}(\bi{v}_{i}\cdot \bi{$\xi$}_{ij})\cos^2\beta_{ij} + f_{\rm e}^{\rm b}(\bi{v}_{i}\cdot \bi{$\xi$}_{ij})\sin^2\beta_{ij}.
	\end{aligned}
	\label{ch2:line-coeffient-tensor-back-force}
\end{equation}
Regarding climb, we consider situations at room temperature where diffusion processes are absent and, hence, climb motion is absent. This is ensured by the simple expedient of setting the viscosity coefficient in glide direction to a very high value, i.e. $B^{\rm c}=1560 ~\upmu {\rm Pa}\cdot {\rm s}$, to ensure negligible mobility in climb direction. 

In summary, for the purpose of calculating the drag force on node $i$, the viscous coefficient tensor and back force vector of the moving segment $_{ij}$ are assembled as 
\begin{equation}
	\begin{aligned}
		\mathcal{B}_{ij}(\bi{v}_{i},\beta_{ij})&=
		\mathcal{B}_{ij}^{\rm g}(\bi{v}_{i},\beta_{ij}) \bi{m}_{ij}\otimes\bi{m}_{ij} +
		\mathcal{B}_{ij}^{\rm l}(\bi{v}_{i},\beta_{ij}) \bi{$\xi$}_{ij}\otimes\bi{$\xi$}_{ij}+
		B^{\rm c}\bi{n}_{ij}\otimes\bi{n}_{ij},
		\\
		\mathcal{F}_{ij}^{\rm b}(\bi{v}_{i},\beta_{ij})&=\mathcal{F}_{ij}^{\rm b,g}(\bi{v}_{i},\beta_{ij})\bi{m}_{ij}+\mathcal{F}_{ij}^{\rm b,l}(\bi{v}_{i},\beta_{ij})\bi{$\xi$}_{ij}.
	\end{aligned}
	\label{ch2:glide-dislocation-coeffient-tensor-back-force}
\end{equation}
For immobile segments, this consideration must be modified as only the expansion and contraction of the dislocation segment need to be considered. The corresponding viscous coefficient tensor and back force are
\begin{equation}
	\begin{aligned}
		\mathcal{B}_{ij}(\bi{v}_{i},\beta_{ij})&=\mathcal{B}_{ij}^{\rm l}(\bi{v}_{i},\beta_{ij})\bi{$\xi$}_{ij}\otimes\bi{$\xi$}_{ij} + B^{\rm c}\left(\bi{m}_{ij}\otimes\bi{m}_{ij}+\bi{n}_{ij}\otimes\bi{n}_{ij}\right),
		\\
		\mathcal{F}_{ij}^{\rm b}(\bi{v}_{i},\beta_{ij})&=\mathcal{F}_{ij}^{\rm b,l}(\bi{v}_{i},\beta_{ij})\bi{$\xi$}_{ij}.
	\end{aligned}
	\label{ch2:physical-junction-dislocation-coeffient-tensor-back-force}	
\end{equation}

Finally, in order to evaluate the total drag force on a given node $i$, the drag force must be integrated over all segments connecting to that node. Making again the simplifying assumption that velocity variations along a segment can be neglected, such that its drag force distributes evenly onto the adjacent nodes, the viscous coefficient tensor function $\mathbcal{B}_i\left(\bi{v}_i\right)$ and the back force vector function $\mathbcal{F}^{\rm b}_i\left(\bi{v}_i\right)$ of the node $i$ are obtained as
\begin{equation}
	\begin{aligned}
		\mathbcal{B}_i\left(\bi{v}_i\right) &= \frac{1}{2}\sum_j \|\bi{l}_{ij}\| \mathcal{B}_{ij}\left(\bi{v}_i,\beta_{ij}\right), \\
		\mathbcal{F}_i^{\rm b}\left(\bi{v}_i\right) &= \frac{1}{2}\sum_j \|\bi{l}_{ij}\| \mathcal{F}_{ij}^{\rm b}\left(\bi{v}_i,\beta_{ij}\right).
	\end{aligned}
	\label{ch2:assemble-nodal-drag-coeffient}
\end{equation}

The velocity $\bi{v}_i$ is now evaluated from the requirement that the net force on each node (the sum of nodal force $\mathbcal{F}_i$ and drag force $\mathbcal{F}_i^{\rm d}$) must be zero. To solve the ensuing nonlinear equation, the Newton iteration method is used. In the $n$th iteration step, the drag force $\bi{F}_i^{\rm d}(n)$ is expressed as
\begin{equation}
	\bi{F}_i^{\rm d}(n)=-\left[\mathbcal{B}_i\left(\bi{v}_i(n)\right)\cdot \bi{v}_i(n)+\mathbcal{F}_i^{\rm b}\left(\bi{v}_i(n)\right)\right].
	\label{ch2:assemble-nodal-drag-frce}
\end{equation}
From this force, the nodal velocity $\bi{v}_i(n+1)$ in the $(n+1)$-th iteration step is then calculated as
\begin{equation}
	\bi{v}_i(n+1) = \bi{v}_i(n)+ \mathbcal{B}_i^{-1}\left(\bi{v}_i(n)\right) \cdot\left[\bi{F}_i+\bi{F}_i^{\rm d}(n)\right],
	\label{ch2:newton-iteration-v-n1}
\end{equation}
where $\mathbcal{B}_i^{-1}\left(\bi{v}_i(n)\right)$ denotes the inverse of the nodal viscosity coefficient tensor.
The iteration can be stopped when the magnitude $\|\bi{F}_i+\bi{F}_i^{\rm d}(n)\|$ of the resultant force  is smaller than the force tolerance $F_{\rm tol}$, which is set equal to $1~{\rm Pa\cdot b^2}$ in this work.

\subsection{Algorithm for decomposition of interface dislocations}
Normally, the interactions of dislocations with TBs are extremely complicated \cite{Zhu-Dislocation–twin-interactions-Acta-2011,Fan-Orientation-influence-Scripta-2015} and a wide range of candidate dislocation configurations need to be considered in order to correctly identify the resulting dislocation configuration. To address this issue, a universal, crystal structure-based interface dislocation decomposition algorithm has been developed.

Suppose an interface dislocation segment of Burgers vector $\bi{b}^{\rm in}$ is to be decomposed. Based on the crystallographic structure of the interface and the adjacent grains, all glide planes parallel to the segment and all Burgers vectors $\left\{\bi{b}_q\right\}$ of glissile dislocations on these planes can be determined. These Burgers vectors serve as basis vectors for the decomposition. Without loss of generality we require $\bi{b}_q\cdot\bi{b}^{\rm in} \ge 0$. On the basis of conservation of Burgers vector, the indefinite equation for the decomposition of $\bi{b}^{\rm in}$ can then be given as
\begin{equation}
	\bi{b}^{\rm in} = \sum_q\left(n_{q} \bi{b}_q\right)+\bi{b}^{\rm r},
	\label{ch2:burg-decomposition}
\end{equation}
where $\bi{b}^{\rm r}$ is the Burgers vector of residual debris left on the interface, and $n_q$ is an integer number of glissile dislocations. Eq. \ref{ch2:burg-decomposition} is an indefinite equation with an infinite number of solutions, however, most of these solutions are energetically very unfavorable. The energy cost of the decomposition can be expressed in terms of the energy barrier factor
\begin{equation}
f=\frac{1}{b^2}\sum_q \left(|n_q|~\|\bi{b}_q\|^2\right)+\|\bi{b}^{\rm r}\|^2-\|\bi{b}^{\rm in}\|^2
\label{ch2:energyfactor}
\end{equation}
where $b$ is the Burgers vector length of a full dislocation. Based on arguments of energy minimization, we impose the following heuristic constraints on the candidates:

(i) The number of dislocations of a given basis vector is constrained to be in the range $-1 \le n_q \le 2$. The asymmetry between positive and negative values reflects the fact that negative $n_q$ represent dislocations with Burgers vectors that are obtuse to the initial dislocation, which is energetically unfavorable. However, such dislocations cannot be completely excluded since experiment \cite{Li-Twinningdislocationmultiplication-AM-2011} shows that energetically unfavorable decomposition modes where $\bi{b}_q\cdot\bi{b}^{\rm in}<0$ may be still induced by stress concentrations. Also, we assume a maximum value for the number of dislocations of the same Burgers vector, which we take to be $n_q \le 2$; 

(ii) The total number of dislocations in the decomposed configuration should be less than a small value, we assume $\sum_q n_q \le 4$; 

(iii) The magnitude of the Burgers vector of the residual debris should be smaller than the Burgers vector of a partial dislocation, $\bi{b}^{\rm p}$, i.e., $\| \bi{b}^{\rm r} \| < \| \bi{b}^{\rm p}\| $; 

(iv) Considering that the reaction barrier cannot be too large, the energy barrier factor $f$ is constrained to be less than 1.

With the given constraints, the indeterminate equation can be solved iteratively to obtain multiple sets of solutions, where the number of Burgers vectors $\bi{b}_q$ in the $s$-th set of Burgers vector combinations is $n_{qs}$ and the Burgers vector of the residual debris is $\bi{b}^{\rm r}_s$.

Next, glide planes are assigned to the dislocations in each set of Burgers vector combinations. 
If the Burgers vector $\bi{b}_q$ can be located on $n_q^{\rm g}$ different glide planes, the number $N_s$ of candidate configurations that can be derived from the $s$-th set of Burgers vector combinations is 
\begin{equation}
	N_s=\prod_q C_{|n_{qs}|+n_q^{\rm g}-1}^{|n_{qs}|}=\prod_q\frac{\left(|n_{qs}|+n_q^{\rm g}-1\right)!}{\left(n_q^{\rm g}-1\right)!|n_{qs}|!}.
	\label{ch2:number-candidate-structures}
\end{equation}
Thus, the total set of candidate dislocation configurations for the decomposition of the Burgers vector $\bi{b}^{\rm i}$ consists of  $\sum_s N_s$ members.

For dislocation-TB reactions, the number of candidate configurations can be further reduced based on a combination of theoretical analysis \cite{Zhu-Dislocation–twin-interactions-Acta-2011} and in situ observations \cite{Wang-Atomic-scale-in-Applied-2009,Wang-Size-dependent-dislocation-twin-Nanoscale-2019}, which lead to the following additional rules:  
(i) Only TDs are allowed to decompose on TBs; 
(ii) Decomposition does not lead to partial dislocations in the lattice; 
(iii) If there exists a candidate configuration without debris, other configurations with debris are eliminated.
(iv) When the energy barrier factor is greater than zero, the reaction stress needs to be increased by an amount proportional to $f$; we evaluate the modified reaction stress as $(1+f)\tau^{\rm r}$.

The decomposition is assumed to occur in two steps: first decomposition on different glide planes, then further decomposition into dislocations with different Burgers vectors.
Based on the assumptions stated above, the final candidate can thus be obtained sequentially by first applying the criterion of maximum local glide stress and then the criterion of smallest possible energy barrier factor. 

For illustration, results of applying the interfacial dislocation decomposition algorithm in simulations of typical dislocation-TB interaction processes where a single dislocation impinges on a TB are shown in \ref{dislocation-TB-interaction}.

\subsection{Computational model}

\begin{figure}[htb]
	\centering
	\includegraphics[width=0.9\linewidth]{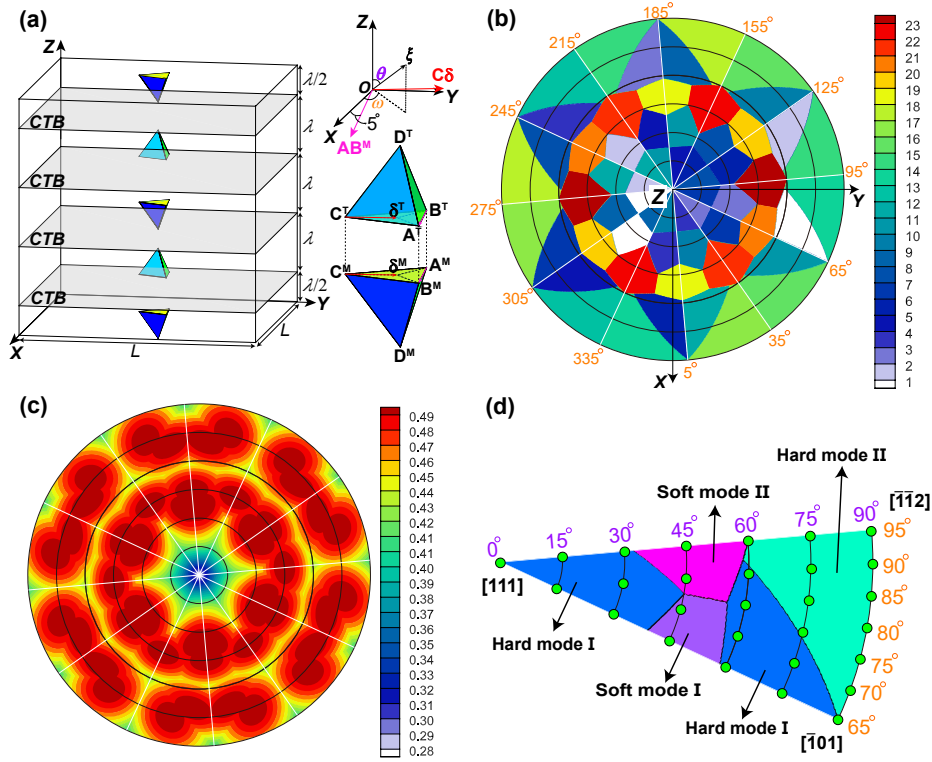}
	\caption{Computational model and polar plots of active slip systems. (a) Geometric model and loading schematics where the Thompson tetrahedra indicate the crystal orientation; the superscripts M and T indicate the matrix and twin, respectively. The polar angle $\theta$ and azimuthal angle $\omega$ define the loading direction $\bm{\xi}$. (b-d) Polar plots of the active slip systems: (b) slip system index (see Table \ref{ch5:tab:modes}), (c) Schmid factor, and (d) activated slip mode. The green circular symbols in (d) indicate the investigated loading directions.}
	\label{ch5:OE-computation-model}
\end{figure}

In individual grains of nanotwinned materials, the twin length (typically of the order of the grain size) is usually more than ten times the thickness of the twinned lamellae \cite{Lu-Plastic-deformation-Acta-2014}. Here we focus on deformation behavior below the grain scale in columnar nanotwinned metals, which leads to a geometrical model where we consider as representative volume element (RVE) a cube of edge length $L$ as shown in Fig. \ref{ch5:OE-computation-model}(a). In this RVE, a multilayer twin structure is introduced where the twin fraction is 1/2, with a twin thickness $\lambda = 160~{\rm nm}$. The TBs are parallel to the $XOY$ plane and equally spaced along the $Z$ direction. Periodic boundary conditions (PBCs) are applied in all spatial directions.
\begin{table}[!htb]
	\caption{Slip systems and slip modes in MTS, where CTB denotes coherent twin boundary, $a$ denotes the lattice constant, the superscripts M and T denote the matrix and twin, respectively.}
	\label{ch5:tab:modes}
	\centering
    \begin{tabular}{llllll}
    \toprule
    Mode  & Index & \multicolumn{2}{l}{Slip   direction}  & \multicolumn{2}{l}{Glide   plane}  \\ \midrule
    \multirow{12}{*}{Hard mode I} & 1     & $\textbf{DB}^{\rm M}$   & $\frac{a}{2}\left[011\right]^{\rm M}$         & $\textbf{ABD}^{\rm M}$   & $\left(\bar{1}\bar{1}1\right)^{\rm M}$ \\
     & 2  & $\textbf{DA}^{\rm M}$ & $\frac{a}{2}\left[101\right]^{\rm M}$ & $\textbf{ABD}^{\rm M}$ & $\left(\bar{1}\bar{1}1\right)^{\rm M}$ \\
     & 3  & $\textbf{DC}^{\rm M}$ & $\frac{a}{2}\left[110\right]^{\rm M}$ & $\textbf{ACD}^{\rm M}$ & $\left(1\bar{1}\bar{1}\right)^{\rm M}$ \\
     & 4  & $\textbf{DA}^{\rm M}$ & $\frac{a}{2}\left[101\right]^{\rm M}$ & $\textbf{ACD}^{\rm M}$ & $\left(1\bar{1}\bar{1}\right)^{\rm M}$ \\
     & 5  & $\textbf{DC}^{\rm M}$ & $\frac{a}{2}\left[110\right]^{\rm M}$ & $\textbf{BCD}^{\rm M}$ & $\left(\bar{1}1\bar{1}\right)^{\rm M}$ \\
     & 6  & $\textbf{DB}^{\rm M}$ & $\frac{a}{2}\left[011\right]^{\rm M}$ & $\textbf{BCD}^{\rm M}$ & $\left(\bar{1}1\bar{1}\right)^{\rm M}$ \\
     & 7  & $\textbf{DB}^{\rm T}$ & $\frac{a}{2}\left[011\right]^{\rm T}$ & $\textbf{ABD}^{\rm T}$ & $\left(\bar{1}\bar{1}1\right)^{\rm T}$ \\
     & 8  & $\textbf{DA}^{\rm T}$ & $\frac{a}{2}\left[101\right]^{\rm T}$ & $\textbf{ABD}^{\rm T}$ & $\left(\bar{1}\bar{1}1\right)^{\rm T}$ \\
     & 9  & $\textbf{DC}^{\rm T}$ & $\frac{a}{2}\left[110\right]^{\rm T}$ & $\textbf{ACD}^{\rm T}$ & $\left(1\bar{1}\bar{1}\right)^{\rm T}$ \\
     & 10 & $\textbf{DA}^{\rm T}$ & $\frac{a}{2}\left[101\right]^{\rm T}$ & $\textbf{ACD}^{\rm T}$ & $\left(1\bar{1}\bar{1}\right)^{\rm T}$ \\
     & 11 & $\textbf{DC}^{\rm T}$ & $\frac{a}{2}\left[110\right]^{\rm T}$ & $\textbf{BCD}^{\rm T}$ & $\left(\bar{1}1\bar{1}\right)^{\rm T}$ \\
     & 12 & $\textbf{DB}^{\rm T}$ & $\frac{a}{2}\left[011\right]^{\rm T}$ & $\textbf{BCD}^{\rm T}$ & $\left(\bar{1}1\bar{1}\right)^{\rm T}$ \\ \midrule
    \multirow{6}{*}{Hard mode II} & 13    & $\textbf{BA}^{\rm M}$   & $\frac{a}{2}\left[1\bar{1}0\right]^{\rm M}$   & $\textbf{ABD}^{\rm M}$   & $\left(\bar{1}\bar{1}1\right)^{\rm M}$ \\
    & 14 & $\textbf{AC}^{\rm M}$ & $\frac{a}{2}\left[01\bar{1}\right]^{\rm M}$ & $\textbf{ACD}^{\rm M}$ & $\left(1\bar{1}\bar{1}\right)^{\rm M}$ \\
    & 15 & $\textbf{BC}^{\rm M}$ & $\frac{a}{2}\left[10\bar{1}\right]^{\rm M}$ & $\textbf{BCD}^{\rm M}$ & $\left(\bar{1}1\bar{1}\right)^{\rm M}$ \\
    & 16 & $\textbf{BA}^{\rm T}$ & $\frac{a}{2}\left[1\bar{1}0\right]^{\rm T}$ & $\textbf{ABD}^{\rm T}$ & $\left(\bar{1}\bar{1}1\right)^{\rm T}$ \\
    & 17 & $\textbf{AC}^{\rm T}$ & $\frac{a}{2}\left[01\bar{1}\right]^{\rm T}$ & $\textbf{ACD}^{\rm T}$ & $\left(1\bar{1}\bar{1}\right)^{\rm T}$ \\
    & 18 & $\textbf{BC}^{\rm T}$ & $\frac{a}{2}\left[10\bar{1}\right]^{\rm T}$ & $\textbf{BCD}^{\rm T}$ & $\left(\bar{1}1\bar{1}\right)^{\rm T}$ \\ \midrule
    \multirow{3}{*}{Soft mode I}  & 19    & $\textbf{BA}^{\rm M/T}$ & $\frac{a}{2}\left[1\bar{1}0\right]^{\rm M/T}$ & $\textbf{ABC}^{\rm M/T}$ & $\left(111\right)^{\rm M/T}$           \\
                                  & 20    & $\textbf{AC}^{\rm M}$ or $\textbf{BC}^{\rm T}$ & $\frac{a}{2}\left[01\bar{1}\right]^{\rm M}$ or $\frac{a}{2}\left[10\bar{1}\right]^{\rm T}$ & $\textbf{ABC}^{\rm M/T}$ & $\left(111\right)^{\rm M/T}$           \\
                                  & 21    & $\textbf{BC}^{\rm M}$ or $\textbf{AC}^{\rm T}$ & $\frac{a}{2}\left[10\bar{1}\right]^{\rm M}$ or $\frac{a}{2}\left[01\bar{1}\right]^{\rm T}$ & $\textbf{ABC}^{\rm M/T}$ & $\left(111\right)^{\rm M/T}$           \\ \midrule
    \multirow{3}{*}{Soft mode II} & 22    & $\textbf{A$\delta$}$    & $\frac{a}{6}\left[\bar{1}2\bar{1}\right]$     & CTB                      & $\left(111\right)$                     \\
                                  & 23    & $\textbf{B$\delta$}$    & $\frac{a}{6}\left[2\bar{1}\bar{1}\right]$     & CTB                      & $\left(111\right)$                     \\
                                  & 24    & $\textbf{C$\delta$}$    & $\frac{a}{6}\left[\bar{1}\bar{1}2\right]$     & CTB                      & $\left(111\right)$    \\ 
                                  \bottomrule                
    \end{tabular}
\end{table}

In Fig. \ref{ch5:OE-computation-model}(a), the orientation of the matrix (M) and the twin (T) are shown as Thompson tetrahedra. The lattice orientation was constructed as follows. Initially. the crystal directions were aligned with the coordinate axes, which are perpendicular to the side surfaces of the simulation cell: $X$ axis $\parallel \textbf{AB}^{\rm M,T}$, $Y$ axis $\parallel \textbf{C}^{\rm M,T}{\textbf{$\delta$}}^{\rm M,T}$, $Z$ axis $\parallel \textbf{D}^{\rm M,T}{\textbf{$\delta$}}^{\rm M,T}$. This is also the geometry considered in \ref{dislocation-TB-interaction}. However, this geometry has the drawback that the glide planes $\textbf{ABD}$ are at the $XZ$ side surfaces connected across the periodic boundary, such that dislocations activated by the same source may meet across the boundary and self-annihilate. This artefact does not occur for the $\textbf{ACD}$ and $\textbf{BCD}$ planes and introduces into the dislocation dynamics an artificial asymmetry  that is a boundary condition artefact. To avoid this problem, the Thomson tetrahedra of the both the matrix and the twins were rotated counterclockwise around the $Z$ axis by a small angle $\omega_{\rm 0}=5^{\circ}$, thus preventing self-annihilation.
 
For loading, a tensile strain rate of $\dot{\varepsilon}=5000 ~{\rm s}^{-1}$ is applied along the 
tension direction $\bi{$\xi$}$ which is defined by the polar angle $\theta$ between the loading axis and the $OZ$ direction, and the azimuthal angle $\omega$ between the projection of the loading axis on the $XOY$-plane and the $OX$-axis, $\bi{$\xi$}=\left(\sin \theta \cos \omega, \sin \theta \sin \omega, \cos\theta \right)$. Material parameters typical of Copper are used in the simulations:
shear modulus $\mu=48~{\rm GPa}$, Poisson's ratio $\nu=0.324$, magnitude of Burgers vector $b=0.256~{\rm nm}$.

In the MTS, the 24 basic slip systems can be divided into four slip modes, corresponding to four dislocation types, as shown in Table \ref{ch5:tab:modes}. Under the assumption that the activation stresses of all slip systems are the same, i.e., isotropic, the polar diagrams of this twin structure have been calculated and plotted by the software Matlab \cite{matlab2022a} and the texture analysis toolkit MTEX \cite{Bachmann-Texture-MTEX-TTP-2010}. Figs. \ref{ch5:OE-computation-model} (b), (c) and (d) indicate for different loading directions $\bi{$\xi$}$ the index of the primary slip system, the corresponding Schmid factor and activation mode, respectively.
It should be noted that the activation mode denotes the type of dislocation mode that operates first when the activation stresses of all slip systems are identical (i.e., isotropic). It does not necessarily represent the slip mode that prevails during deformation. 
It can be seen that the Schmid factor polar diagram and the activation mode polar diagram both vary periodically along the loading azimuth and are symmetric. These symmetries can be exploited by restricting the investigation to one sector, which we take to be the azimuthal range $65^{\circ} \le \omega \le 95^{\circ}$ in the polar diagram. 28 loading directions in this sector (green dots in Fig. \ref{ch5:OE-computation-model}(d)) were selected to implement DDD simulations for investigating the effect of loading axis orientation on the deformation behavior of the MTS.

To ensure the physical fidelity of the simulations, the initial dislocation density $\rho_{0}$ was determined based on experimental evidence. Transmission electron microscopy (TEM) utilizing two-beam diffraction techniques has shown that for nanotwinned copper dominated by Hard Mode II, the dislocation density reaches approximately $10^{13}$ to $10^{14} ~{\rm m}^{-2}$ at 1\% strain \cite{Cheng-Extra-strengthening-Science-2018}. 
Balancing the computational cost of DDD with the necessity of maintaining a statistically representative defect population, we established an initial dislocation density of $5\times 10^{13} ~{\rm m}^{-2}$. This value lies within the experimentally observed range, ensuring sufficient dislocation interactions without introducing excessive computational overhead.

The choice of the dislocation multiplication mechanism in nanotwinned metals warrants careful justification. MD simulations \cite{You-Plastic-anisotropy-Acta-2013,Cheng-Extra-strengthening-Science-2018} indicate that for extremely fine microstructures, dislocations predominantly originate from interfaces: Hard Mode I dislocations nucleate directly from TBs, while other active modes typically initiate from GB or GB-TB intersections. 
Theoretical analysis by Dao et al. \cite{Dao_QuantitativeUnderstandingMechanical_ActaMater._2007} further supports this, suggesting that for grain sizes below 100 nm, the stress required to activate intragranular Frank-Read (FR) sources would be three times higher than experimental measurements, implying that boundary-assisted nucleation is essential at the finest scales. Experimental observations of pre-deformed samples \cite{You-Tensile-behavior-Acta-2011} indicate that dislocation storage within the twinned lamellae was initially negligible, suggesting the absence of a pre-existing intragranular dislocation network, again pointing to nucleation at interfaces as main multiplication mechanism. 

In the situations that are in the focus of the present study, the macroscopic yield strength in this range follows a smooth Hall-Petch relation without inflection points \cite{You-Tensile-behavior-Acta-2011}, indicating that the strength is governed by the geometric confinement of slip rather than the specific mechanism of nucleation. To capture the physics of confined slip while minimizing the number of phenomenological parameters associated with boundary nucleation, we adopted a simplified model using pre-existing Frank-Read sources distributed within the twin lamellae while ensuring that the sources are sufficiently long to ensure that slip activation is not source limited. This is achieved by considering sources with source length between $\lambda$ and $3 \lambda$ that are, when located on hard mode I and II slip systems, are oriented such that they thread across the twin lamellae. This avoids artefacts due to the artificially introduced pinning points acting as strong obstacles in the central region of the twin lamellae. 

In the final step of the process of setting the dislocation sources, to account for the distinct lattice resistance of different dislocation types, the lengths of twinning sources were scaled by a factor of $1/\sqrt{3}$ relative to full dislocations (since the magnitude of Burgers vector for TD $b_{\rm t} = b/\sqrt{3}$). This scaling ensures isotropic activation criteria where the activation stress $\tau_{\rm a} \propto \mu b / L_{\rm FR}$ remains consistent across different modes.
The initialized structures were relaxed for 5,000 time steps ($\approx$ 25 ns), which was found sufficient to ensure that the residual plastic activity in the relaxed structure is negligible. Four statistical realizations of the initial conditions were simulated for each loading orientation to ensure statistical robustness.

To ensure that the RVE can adequately capture the dislocation processes in nanotwinned grains, artefacts of the PBCs need to be minimized by making the simulated volume large enough. This implies that the RVE size $L$ needs to be several times larger than the characteristic microstructure size. In DDD simulations of bulk materials, the characteristic size is the mean dislocation spacing, and the RVE size should be no less than four times the mean dislocation spacing \cite{Fan-Strain-rate-Nat-Commun-2021}. However, in MTS, the characteristic size is not only related to the dislocation spacing, but also to the dislocation-TB microstructure size, so the characteristic size is different for different slip modes. A size scaling study of our DDD simulations indicates that, if the loading axis is not parallel to the TB ($\theta \le 75^{\circ}$), the four-layer twin structure shows reproducible behavior and stable flow stress in the plastic stage, so $L=4\lambda$ is used. If the loading axis is parallel to TBs ($\theta = 90^{\circ}$), hard mode II dislocations may transmit TBs and form a microstructure that encompasses several twin layers. This leads to increased fluctuations in the mechanical behavior between different realizations of the initial conditions. Therefore, in this case the system size is increased to $L=6\lambda$ to ensure consistent results.

\section{Results and Discussion}\label{results-discussions}

\subsection{Mechanical behavior and dislocation density evolution}
\begin{figure}[htb]
	\centering
	\includegraphics[width=0.85\linewidth]{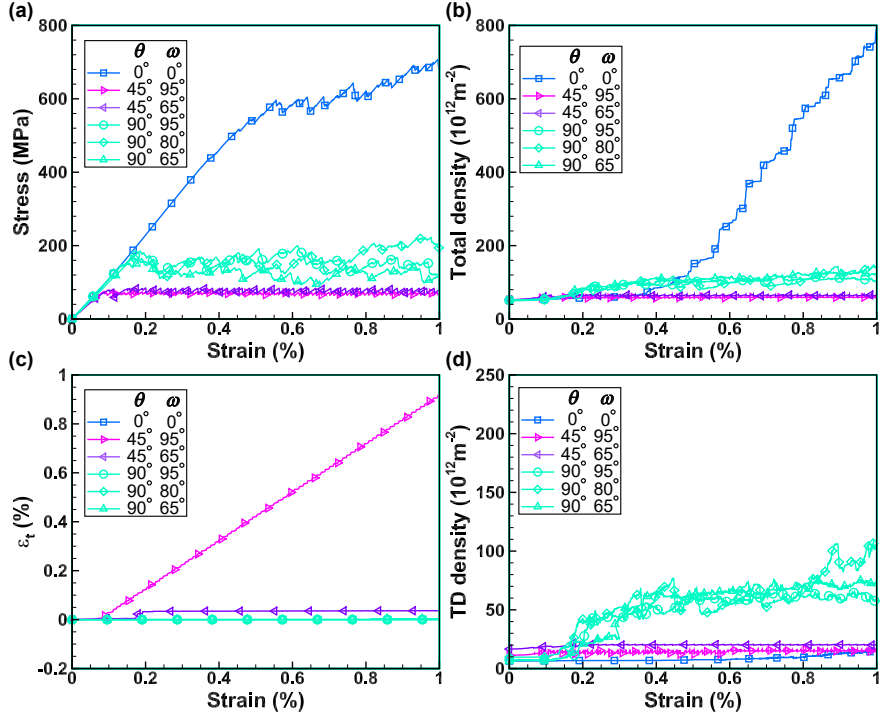}
	\caption{(a) stress, (b) total dislocation density, (c) twinning strain and (d) TD density versus loading strain for various loading orientations. The color scheme is the same as in Fig. \ref{ch5:OE-computation-model}(d). The curves of $\theta=0^{\circ}$ and $90^{\circ}$ in (c)  overlap, in both cases the twinning strains are zero.}
	\label{ch5:OE-ss-curves}
\end{figure}

Fig. \ref{ch5:OE-ss-curves} shows stress-strain response, evolution of the total dislocation density, twinning strain and TD density evolution for six typical orientations of the loading axis. It can be seen that the trends of mechanical response and dislocation density evolution in MTS are qualitatively similar to the findings reported by Wei et al. for bicrystalline twinned micropillars \cite{Wei-Effects-of-Acta-2019}, and also consistent with the trends reported in experimental work on orientation effects in columnar nanotwinned copper \cite{You-Plastic-anisotropy-Acta-2013}.

Typical results for situations where the loading axis is perpendicular to the TBs, i.e. $\theta=0^{\circ}$, are shown by the blue curves in Fig. \ref{ch5:OE-ss-curves}. In this case, the flow stress in the plastic stage reaches a quite high value of about $550\sim 700~{\rm MPa}$. After the onset of plastic flow, one observes pronounced strain hardening as the total dislocation density rises to $8\times10^{14}~{\rm m}^{-2}$ within about 0.5\% plastic strain. At the same time, almost no twinning deformation occurs and the TD density remains unchanged.

Typical results for situations where the loading axis is inclined to the TB, namely $15^{\circ}\le \theta \le 75^{\circ}$, are indicated by the pink and purple curves in Fig. \ref{ch5:OE-ss-curves}, representing an inclination angle of $\theta=45^{\circ}$ and two different azimuth angles. 
In these orientations, the MTSs show little strain hardening and negligible changes in dislocation density. As demonstrated in Fig. \ref{ch5:OE-computation-model}(d), despite the fact that both hard-mode dislocations can be activated in these orientations (with largest Schmid factors), owing to the presence of TBs which inhibit multiplication of hard-mode dislocations, plastic deformation is instead dominated by the soft modes.  

For cases represented by the pink curve in Fig. \ref{ch5:OE-ss-curves}(c) ($\omega=95^{\circ}$), the maximum shear stress direction is almost parallel to the Burgers vector of TDs (see  Fig. \ref{ch5:OE-computation-model}(d) and Table \ref{ch5:tab:modes}, where they are identified as $\textbf{C$\delta$}$). In this case, plastic deformation occurs almost exclusively by TD slip. As a consequence, the twinning strain accounts increases linearly with total strain, and this increase accounts almost fully for the total strain increase. In a number of simulations, as illustrated by the purple curve in Fig. \ref{ch5:OE-ss-curves}(c), deformation is dominated by lattice dislocations parallel to the TBs ($\textbf{BC}^{\rm M}$ and $\textbf{AC}^{\rm T}$ in Table \ref{ch5:tab:modes}) while twinning deformation is practically absent.

When the loading axis is parallel or near-parallel to the TB, typical results correspond to the three green curves shown for $\theta=90^{\circ}$ in Fig. \ref{ch5:OE-ss-curves}. In these orientations, the stress-strain curves are characterized by a pronounced stress drop immediately after the onset of plastic flow. This stress drop is accompanied by a marked increase in TD density. This indicates that the stress drop is associated with the decomposition and/or transmission of hard mode II dislocations at the TBs which occurs once the critical reaction stress is reached. These reactions lead to an increase in TD density and, as the constraining effect of the TBs on dislocation source activation is reduced, to a reduced flow stress.

\begin{figure}[htb]
	\centering
	\includegraphics[width=0.8\linewidth]{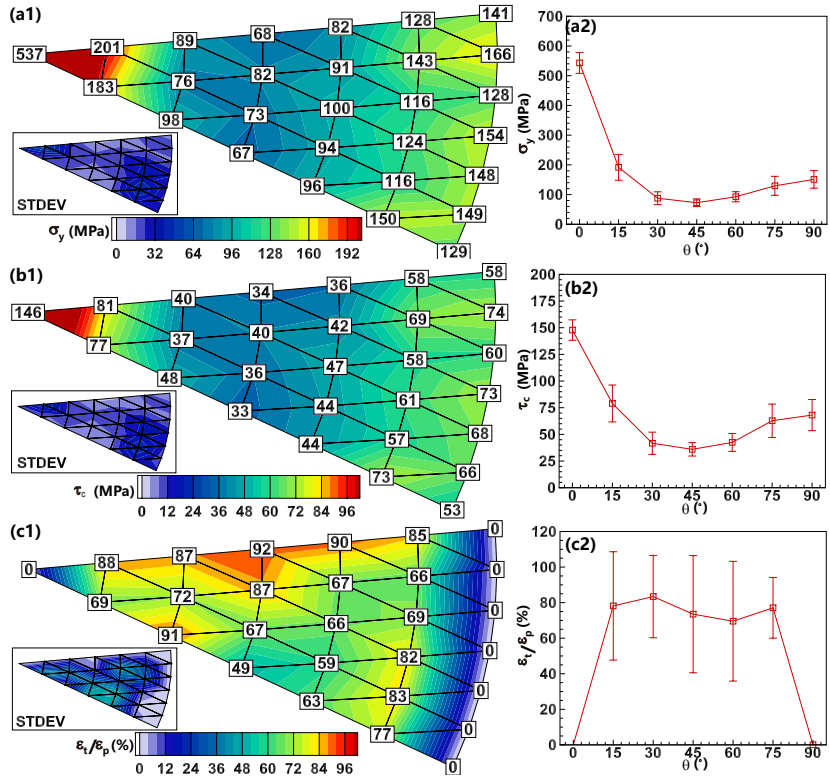}
	\setlength{\belowcaptionskip}{0pt}
	\setlength{\abovedisplayskip}{-3pt}
	\caption{Dependence on loading axis orientation of (a) yield stress $\sigma_{\rm y}$, (b) CRSS $\tau_{\rm c}$ and (c) ratio $\varepsilon_{\rm t}/\varepsilon_{\rm p}$ between twinning strain and total plastic strain; all quantities are determined at $\varepsilon_{\rm p}=0.2\%$; left: polar plots, (right) dependency on pole angle. The embedded polar plots and error bars indicate sample standard deviations (STDEV).}
	\label{ch5:OE-polar-figures}
\end{figure}

Fig. \ref{ch5:OE-polar-figures}(left) shows pole figures of the averaged values of yield stress ($\sigma_{\rm y}$, defined as stress at $0.2\%$ plastic strain), CRSS ($\tau_{\rm c} =\sigma_{\rm y}m_{\rm a}$, with $m_{\rm a}$ as the Schmid factor of the active slip system), and twinning strain percentage of the plastic strain ($\varepsilon_{\rm t}/\varepsilon_{\rm p}$), all plotted as functions of the loading axis orientation. The curves on the right-hand side of Fig. \ref{ch5:OE-polar-figures} illustrate the variation of the same physical quantity as a function of polar angle, with different azimuths of the same polar angle averaged to the same point. The embedded polar plots and error bars indicate sample standard deviations (STDEV). In general, strength and degree of twinning deformation are primarily controlled by the polar angle and show a much less pronounced dependence on the azimuth angle. 

When the loading axis is perpendicular to the TBs, there is no twinning deformation while both yield stress and CRSS reach global maxima of 535 MPa and 146 MPa, respectively. 
Comparison with Fig. \ref{ch5:OE-computation-model}(d) indicates that, in this regime, deformation is controlled by hard mode I dislocations. 

As the polar angle is increased from $15^{\circ}$ to $75^{\circ}$, yield stress and CRSS maintain first decrease, then pass through a global minimum when $\theta$ is near $45^\circ$, and then rise again. Near the CRSS minimum, twinning accounts for more than half of the total  strain, though the precise amount of twinning depends in a complex manner on the azimuthal angle (Fig. \ref{ch5:OE-polar-figures}(c1)). 

Comparison with Fig. \ref{ch5:OE-computation-model}(d) shows that deformation in soft mode II prevails across a wide range of polar angles, $15^{\circ} \le \theta \le 75^{\circ}$, irrespective of which slip system has the highest Schmid factor. 

As the polar angle $\theta$ further increases towards $90^\circ$, twinning deformation again becomes irrelevant and the yield stress and CRSS increase to 160 MPa and 73 MPa, respectively. 
Comparison with Fig. \ref{ch5:OE-computation-model}(d) indicates that hard mode II dislocations operate and dominate plastic deformation in this regime. Besides, when $\theta=90^\circ$, one observes a slight dependence of yield stress and CRSS on azimuth angle; these quantities attain a local maximum  at azimuth angles near $\omega=80^\circ$. Comparing with the Schmid factors shown in Fig. \ref{ch5:OE-computation-model}, it can be inferred that the dependence on azimuth angle violates Schmid's law according to which yield stress and CRSS increase with higher Schmid factor. 

\subsection{Slip mode and dislocation microstructure evolution of MTS}

\begin{figure}[htb]
	\centering
	\includegraphics[width=0.8\linewidth]{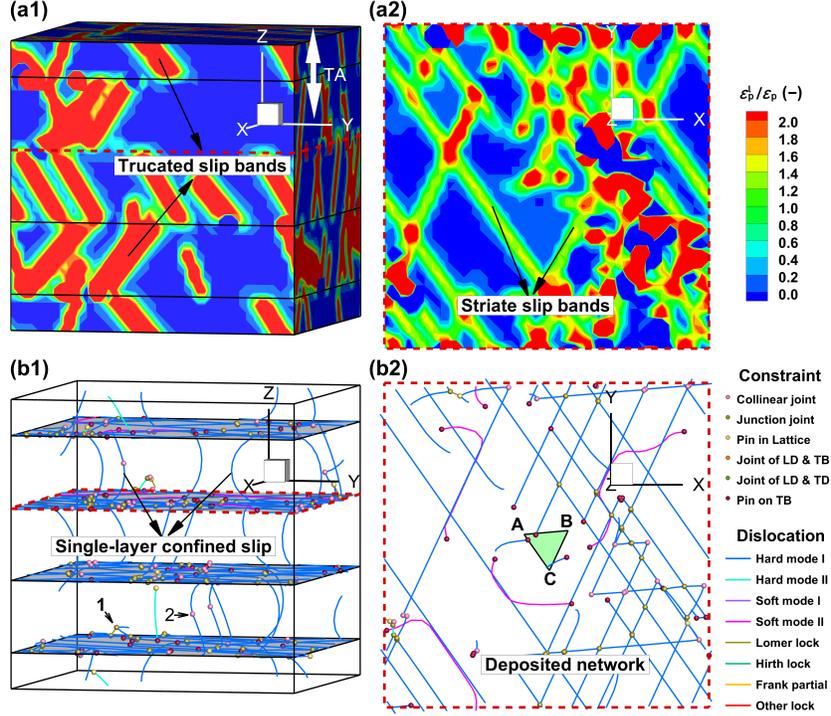}
	\caption{Strain distribution and dislocation microstructure at $\varepsilon=0.53\%$, $\bm{\xi} = \left[111\right]$: (a) slip band nephograms ($\varepsilon_{\rm p}^{\rm l}/\varepsilon_{\rm p}$) and (b) dislocation microstructures; (a1) and (b1) present overall views of the multilayer twin structure, (a2) and (b2) present section views along the TB marked by the red dashed line in the overall view where (b2) represents a layer of thickness 100nm around the TB; the bidirectional arrow in (a1) indicates the tension axis (TA), the numbered arrows in (b1) point to typical dislocation interactions, a legend of the numbers is given in Table \ref{ch5:tab:dislocation-interactions}.}
	\label{ch5:SB-DD-111}
\end{figure}

\begin{figure}[htb]
	\centering
	\includegraphics[width=0.90\linewidth]{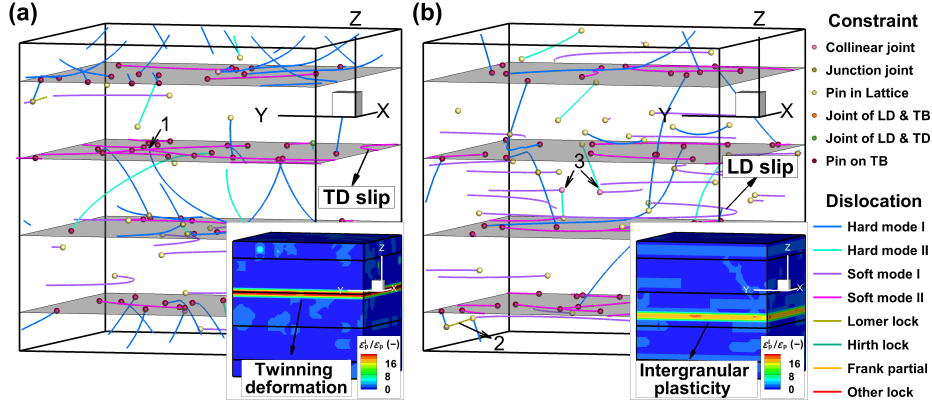}
	\caption{Strain distribution and dislocation microstructures at $\varepsilon=1\%$, $\theta=45^{\circ}$: (a) $\omega=95^{\circ}$, (b) $\omega=65^{\circ}$; the numbered arrows refer to typical dislocation intereactions, a legend of the numbers is given in Table 2; the insets show slip band nephograms illustrating the strain distribution.}
	\label{ch5:SB-DD-45}
\end{figure}

\begin{figure}[htb]
	\centering
	\includegraphics[width=0.85\linewidth]{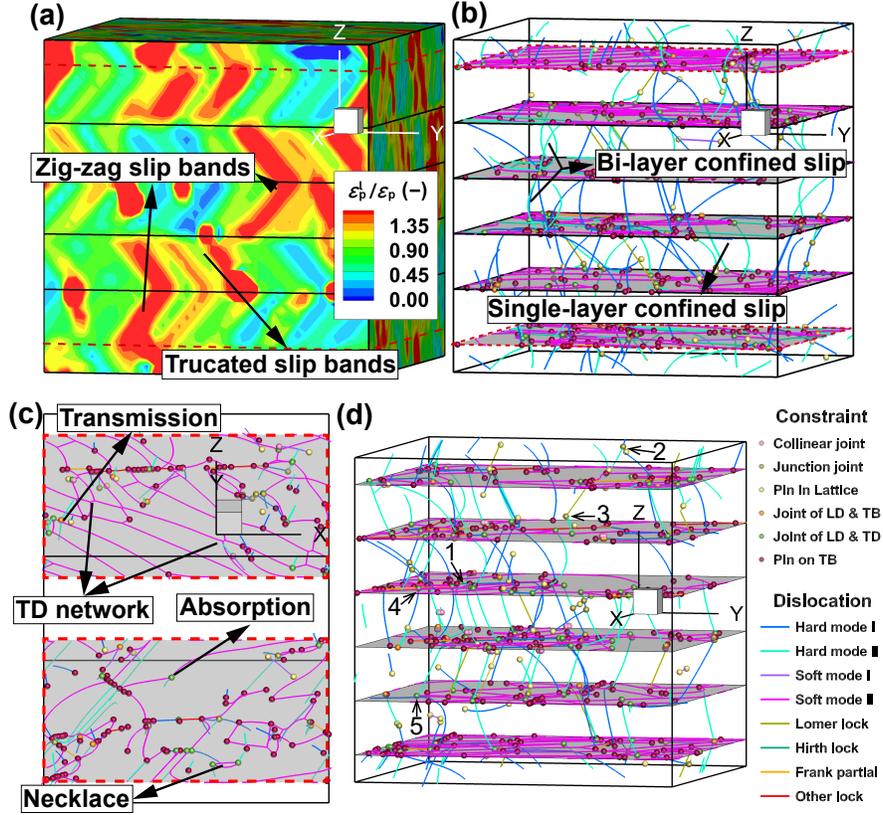}
	\caption{Strain distribution and dislocation microstructures for samples loaded parallel to the TBs ($\theta = 90^{\circ}$); (a-c) slip band nephograms, dislocation microstructure and dislocation configurations on TBs, all for $\omega=95^{\circ}$ and $\varepsilon=0.64\%$; (d) dislocation microstructure  showing dislocation interactions in the sample with axis orientation $\omega=65^{\circ}$ at strain $\varepsilon=1\%$.}
	\label{ch5:SB-DD-90}
\end{figure}

\cref{ch5:SB-DD-111,ch5:SB-DD-45,ch5:SB-DD-90} show slip band nephograms and dislocation microstructures for the previously discussed three typical loading orientations. In them, slip localization is described in terms of the ratio of local plastic strain to global plastic strain, $\varepsilon_{\rm p}^{\rm l}/\varepsilon_{\rm p}$. The RVE is meshed and divided into 41 subcells along all three directions, and dislocation slip is  averaged over the subcells to obtain local plastic strain.
Table \ref{ch5:tab:dislocation-interactions} provides a list of typical dislocation reactions and gives the locations (marked by numbered arrows in the figures) where these reactions can be seen in the figures. 

\begin{table}[htb]
	\caption{Common dislocation-dislocation interactions in MTSs and their locations, where HI, HII, SI and SII are abbreviations for hard mode I, hard mode II, soft mode I and soft mode II dislocations, respectively.}
	\label{ch5:tab:dislocation-interactions}
	\centering
    \begin{tabular}{lllll}
        \toprule
        \multicolumn{2}{l}{Interactants}& Typical interaction equation  & Interaction product  & Location \\ \midrule
        HI &   HI            & $\textbf{AD}\left(\textbf{ACD}\right)+\textbf{DB}\left(\textbf{BCD}\right) \rightarrow \textbf{AB}$                                   & Lomer lock            & Fig. \ref{ch5:SB-DD-111}-1	\\
        HI &   HI            & $\textbf{AD}\left(\textbf{ACD}\right)+\textbf{DA}\left(\textbf{ABD}\right) \rightarrow \textbf{0}$                                    & Collinear junction    & Fig. \ref{ch5:SB-DD-111}-2 \\
        SII&   SII         & $\textbf{A$\delta$}\left({\rm CTB}\right)+\textbf{C$\delta$}\left({\rm CTB}\right) \rightarrow \textbf{$\delta$ B}\left({\rm CTB}\right)$& TD (glissile partial) &Fig. \ref{ch5:SB-DD-45}-1 \\
        SI &   HI         &  $\textbf{DA}\left(\textbf{ACD}\right)+\textbf{AB}\left(\textbf{ABC}\right) \rightarrow \textbf{DB}$  & Lomer lock &Fig. \ref{ch5:SB-DD-45}-2 \\ 
        SI &   HII          & $\textbf{AC}\left(\textbf{ACD}\right)+\textbf{CA}\left(\textbf{ABC}\right) \rightarrow \textbf{0}$                                    & Collinear junction    & Fig. \ref{ch5:SB-DD-45}-3	\\
        HI &   HII          & $\textbf{DA}\left(\textbf{ACD}\right)+\textbf{BC}\left(\textbf{BCD}\right) \rightarrow \textbf{DB}/\textbf{AC}$                       & Hirth lock            & Fig. \ref{ch5:SB-DD-90}-1	\\
        HI &   HII          & $\textbf{DA}\left(\textbf{ABD}\right)+\textbf{AC}\left(\textbf{ACD}\right) \rightarrow \textbf{DC}\left(\textbf{ACD}\right)$          & Glissile junction     & Fig. \ref{ch5:SB-DD-90}-2	\\
        HII&   HII         & $\textbf{CA}\left(\textbf{ACD}\right)+\textbf{AB}\left(\textbf{ABD}\right) \rightarrow \textbf{CB}$                                   & Lomer lock            &  Fig. \ref{ch5:SB-DD-90}-3	\\
        HI &   SII          & $\textbf{DA}\left(\textbf{ACD}\right)+\textbf{A$\delta$}\left({\rm CTB}\right) \rightarrow \textbf{D$\delta$}$                          & Frank partial     & Fig. \ref{ch5:SB-DD-90}-4   \\
        HII&   SII         & $\textbf{AC}\left(\textbf{ACD}\right)+\textbf{C$\delta$}\left({\rm CTB}\right) \rightarrow \textbf{A$\delta$}\left({\rm CTB}\right)$                          & TD (glissile partial) &  	Fig. \ref{ch5:SB-DD-90}-5 \\ \bottomrule
    \end{tabular}
\end{table}

When the loading axis is perpendicular to the TBs ($\bm{\xi}=\left[111\right]$) as shown in Fig. \ref{ch5:SB-DD-111}(a), slip bands forming within the twin layers are truncated by the TBs. At the same time, there is no indication of significant dislocation glide on TBs. According to Fig. \ref{ch5:OE-computation-model}(d), the slip systems activated under this loading orientation belong to hard mode I. The dislocations are of non-screw type when they impinge on TBs, and their transmission or decomposition at the TB require substantial shear stresses \cite{Jin-Interactions-between-Acta-2008}. This explains the hairpin structure resulting from confined layer slip (CLS) of hard mode I dislocations moving within single twin layers (Fig. \ref{ch5:SB-DD-111}(b1)), and the formation of a dense network of disclocations deposited on both sides of the TBs (Fig. \ref{ch5:SB-DD-111}(b2)). Since the deposited dislocations cannot glide on the TBs, the inter-dislocation angle is equal to the $60^\circ$ angle between the triangular edges of the Thompson tetrahedron and the network corresponds to streaks on the TB profile. The deposited dislocations lead to an increase in dislocation density; at the same time their back stress impedes the glide of hairpin dislocations within the layers and causes work hardening. 
In addition, reactions between dislocations within the layers also contribute to hardening, as can be seen from the features indicated by arrows 1 and 2 in Fig. \ref{ch5:SB-DD-111}(b1). 
These features correspond to a Lomer lock and a collinear annihilation junction, respectively, which arise as reaction products between hard mode I dislocations of different slip systems as shown in lines 2-3 of Table \ref{ch5:tab:dislocation-interactions}. 
Lomer locks may deposit in front of TBs and impede glide of hard mode I dislocations in their surrounding, while collinear annihilation joints may undergo confined slip within the twin layer together with two hard mode I dislocations from two different slip planes; the resulting strong kinematic constraints again are a cause of hardening. Moreover, hard mode I dislocations deposited on both sides of the TB cause back stresses that also impede the motion of hard mode I dislocations within the twin layer. In summary, hard mode I dislocations mutually interact both via direct reaction and through back stresses, both mechanisms acting in conjunction to cause significant strain hardening.

The dislocation microstructure shown in Fig. \ref{ch5:SB-DD-111} is in qualitative agreement with observations in similar layered structures under comparable loading orientations \cite{Misra-Length-scale-dependent-deformation-Acta-2005,Lu-Dependence-dislocation-Acta-2017,Lu-Orientation-and-IOP-2017}. 
Some literature sources report observations according to which hard mode I dislocations may thread across multilayer TBs and form long chained dislocations, as observed in an experiment on columnar nanotwinned materials \cite{Lu-Dependence-dislocation-Acta-2017}. The present DDD simulations cannot capture this feature, since simple geometrical analysis shows that, for a hard mode I dislocation transmitted across a TB under loading perpendicular to the TB, the external shear stress acting on the outgoing dislocation would be negative, which makes transmission highly unlikely. This observation poses a conundrum how the threading dislocations are formed. Here we can merely state that their formation by sequential glide-cum-transmission events, as envisaged by Lu et. al. \cite{Lu-Dependence-dislocation-Acta-2017}, is unlikely because the geometry does not allow for such motion. 

One may also note some differences between the simulation results in our paper and findings from large-scale MD simulations considering comparable geometries  \cite{You-Plastic-anisotropy-Acta-2013,Li-Dislocation-nucleation-Nature-2010}. 
The plastic flow stage of such MD simulations is characterized by massive dislocation nucleation at TBs, whereas no confinement of slip of hard mode I dislocations within the layers is observed. 
There are two simple reasons for these differences: 
First, in MD simulations, strain rates are tremendously high and accordingly the stress levels are much higher than in DDD simulations or in standard experiments.
Second, in the mentioned MD simulations the initial structure is dislocation free, in particular there are no sources. 
As a consequence, at the high stress levels and in absence of pre-existing dislocations deformation is accomodated by massive dislocation nucleation which typically leads to a giant stress drop after yield.
At the same time, the high stress levels render TBs ineffectual as dislocation obstacles. 
These features and the ensuing dislocation structures may, in fact, be artefacts of the dislocation-free initial conditions and high strain rates since, in real crystals, initial dislocations are always present and deformation rates of the order of $10^8$/s, as typical of MD simulations, have never been used in experiments on nanotwinned materials. 

When the loading axis is inclined to the TBs, depending on azimuth angle we find two different types of dislocation microstructures and slip distributions as shown in Fig. \ref{ch5:SB-DD-45}. 
For most inclined orientations, simulation results are similar to Fig. \ref{ch5:SB-DD-45}(a) ($\theta,\omega) = (45^{\circ},95^{\circ}$, the evolutions with strain correspond to the pink curves in Fig. \ref{ch5:OE-ss-curves}). 
In this case deformation occurs predominantly by activation and reactions of soft mode II dislocations on TBs, and accordingly plastic strain localizes on the TBs in the form of twinning deformation. 
Since reactions of TDs of two different Burgers vectors generate mobile TDs, no significant hardening effect is observed. 
This behavior is consistent with experimental observations on microstructures with a twin thickness below 200 nm \cite{Lu-Orientation-and-IOP-2017}. 
However, when the Schmid factor of TDs is lower than for soft mode I dislocations, deformation may be governed by the activation and glide of soft mode I dislocations as depicted in Fig. \ref{ch5:SB-DD-45}(b) ($\theta=45^{\circ},~\omega =65^{\circ}$, corresponding to the purple curves in Fig. \ref{ch5:OE-ss-curves}). 
This leads to dislocation motion within the twin layers, accompanied by forest-type dislocation-dislocation reactions in the layers, as indicated by the arrows pointing to products of typical reactions listed in lines 4-6 of Table \ref{ch5:tab:dislocation-interactions}. These reactions between soft mode I dislocations and both hard mode I and II dislocations lead to formation of Lomer locks and collinear joints, respectively, which causes some hardening. Therefore, when the loading axis is inclined to the TB, soft mode II slip, which can proceed without hardening, may ultimately dominate deformation even in situations where soft mode I dislocations experience a higher resolved shear stress.

By thoroughly examining the simulation results mentioned above, along with the insights provided in Fig. \ref{ch5:OE-computation-model}(d), it is evident that certain inclined orientations of the loading axis lead to initial activation of the two hard modes. 
However, the existence of TBs impedes the movement of these hard mode dislocations, ultimately leading to a deformation primarily sustained by the aforementioned two soft mode dislocation types.

When the loading axis is parallel to the TBs ($\theta=90^{\circ}$), irrespective of azimuth angle the resulting strain distribution is similar to that shown for $\omega=95^{\circ}$ in Fig. \ref{ch5:SB-DD-90}(a). In these orientations, slip bands exhibit a zig-zag shape which crosses one or two TBs, while a few slip bands are truncated by TBs and thus remain confined to single layers of the MTS. The dislocation microstructure does not directly match the slip band pattern. Most hard mode II and some hard mode I dislocations experience CLS where they remain constrained within a single layer of the twinned structure, with only a few hard mode II dislocations being able to transmit across TBs and to glide in two or more twin layers. Comparison with the slip distribution indicates that these unconstrained dislocations account for a significant fraction of the plastic strain. This is consistent with microstructural observations \cite{You-Tensile-behavior-Acta-2011,Cheng_UnravelingOriginExtra_ProceedingsoftheNationalAcademyofSciences_2022} and MD simulations \cite{Bu_TranstwinDislocationsNanotwinned_ScriptaMaterialia_2023}, which indicate that hard mode II dislocations that transmit across TBs can glide under reduced stress to produce massive plastic deformation, forming zig-zag shaped slip bands.

The distribution of TDs on TBs is illustrated by the section in Fig. \ref{ch5:SB-DD-90}(c). Four typical configurations related to TD can be distinguished:
\begin{enumerate}
	\item Absorption: hard mode II dislocations are absorbed by TBs and decompose into TDs, so that the parts of the hard mode II dislocation within the twin layers are confined by the TBs.
	\item Transmission: Hard Mode II dislocations are transmitted across TBs and break the limits of monolayer confined slip, generating a large amount of plasticity at reduced shear stress.
	\item Necklace: Two absorbed dislocations with the same glide direction annihilate at the TB, leaving a small ring of TDs, to which two hard mode II dislocations in the adjacent layers are attached and glide in concert. This necklace configuration is similar to the transmitted configuration and therefore its presence can cause softening. In addition, it has been argued that this configuration helps to maintain the stability of the TBs and enhance reversibility during cyclic loading \cite{Pan-History-independent-cyclic-Nature-2017}.
	\item TD network: absorbed configurations crossing each other along the glide direction react to form mobile TDs (with typical reactions as represented in line 4 of Table \ref{ch5:tab:dislocation-interactions}). These TDs of varied Burgers vectors constitute a network.
\end{enumerate}

In addition to hard mode II dislocation reactions at TBs, the activation of hard mode I or II dislocations can lead to extensive dislocation reactions within the MTS layers. As shown in Fig. \ref{ch5:SB-DD-90}(d), when $\theta=90^{\circ}$, there are the last 5 reactions involving hard mode I and II dislocations as listed in Table \ref{ch5:tab:dislocation-interactions}. 

In twin layers, hard mode I and II dislocations may react with each other to generate Hirth locks or glissile joints (arrows 1 and 2 in Fig. \ref{ch5:SB-DD-90}(d)). 
Reaction between hard mode II dislocations may lead to the formation of Lomer locks (arrow 3 in Fig. \ref{ch5:SB-DD-90}(d)). In  addition, hard mode I and II dislocations may react with soft mode II dislocations to create immobile Frank partial dislocations and mobile TDs, respectively (arrows 4 and 5 in Fig. \ref{ch5:SB-DD-90}(d)).

\subsection{Model for prediction of orientation effects in MTSs}

As demonstrated by the analysis above, the presence of TBs introduces an anisotropy of the CRSS in MTS. To deal with the anisotropy, we use different CRSS values in Schmid's law when applying it to  predict the yield stress as a function of loading axis orientation by setting 
\begin{equation}
	\sigma_{\rm y}^{\rm p}={\rm min}_s\left(\frac{\tau_{\rm c}^{s}}{m_s}\right),
	\label{ch5:sigmayp}
\end{equation}
where $s$ labels the different types of slip systems and $\tau_{\rm c}^s$ represents the corresponding CRSS. Our previous analysis indicates that, for the loading axis vertical to the TB, dislocations of hard mode I dominate the yield process. For this deformation mode, the CRSS was determined above as $\tau_{\rm c}^{\rm HI}=148~{\rm MPa}$. In case of a loading axis parallel to the TB, hard mode II dislocation interactions with TBs control the yield process. The CRSS of $\theta=90^{\circ}$ is supposed to be the CRSS of hard mode II, which we estimate from Fig. \ref{ch5:OE-polar-figures}(a) as $\tau_{\rm c}^{\rm HII}=68~{\rm MPa}$. Finally, if the loading axis is inclined to the TB, soft mode dislocations are activated and dominate the yield process, thus the result for $\theta=45^{\circ}$ is chosen as the CRSS of soft mode I and II, namely $\tau_{\rm c}^{\rm SI}=\tau_{\rm c}^{\rm SII}=36~{\rm MPa}$.

\begin{figure}[htb]
	\centering
	\includegraphics[width=1\linewidth]{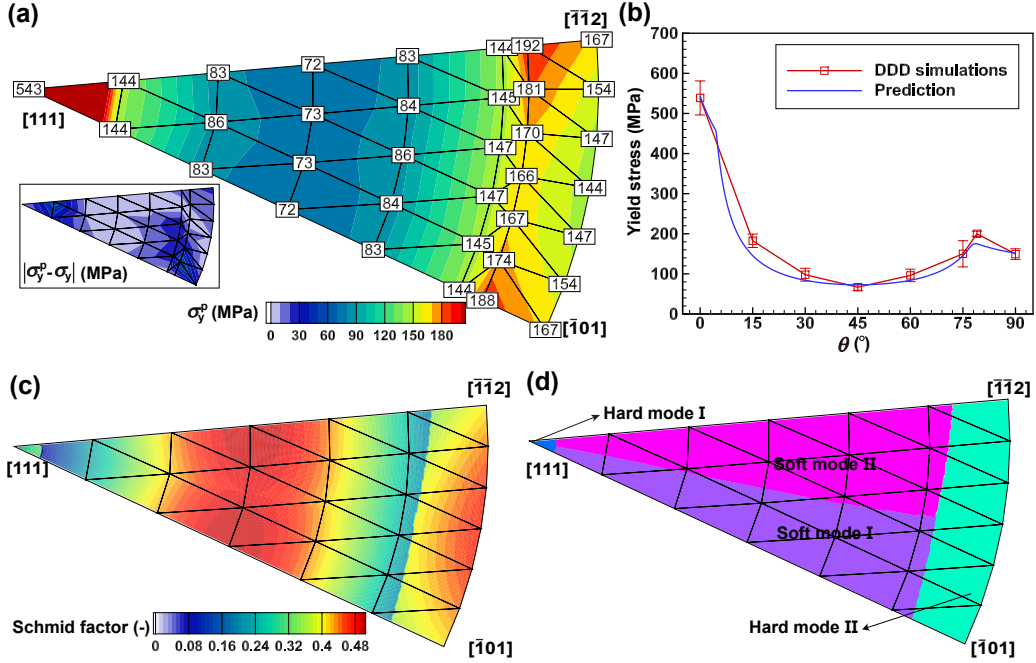}
	\caption{
	Predicted of orientation deendence of the yield stress of MTS. (a) polar figures of predicted yield stress and absolute error (inset); (b) predicted and simulated yield stress (average over azimuth angle) versus polar angle $\theta$ of the loading axis; (c) polar plot of predicted Schmid factor for the operating slip system; (d) pole figure for predicted slip mode.}
	\label{ch5:OE-Prediction}
\end{figure}

In the analytical calculation, the model predicts a distinct local maximum in yield strength at a loading polar angle of $\theta \approx 79^\circ$. 
It is worth noting that this specific feature has not been reported in previous molecular dynamics simulations \cite{You-Plastic-anisotropy-Acta-2013}.
To verify this prediction, supplementary DDD simulations were performed for 7 equidistant azimuth angles ($\omega$) specifically fixed at $\theta=79^{\circ}$. 
The comparisons between model predictions and simulation results are presented in Fig. \ref{ch5:OE-Prediction}.

The polar plots in Figs. \ref{ch5:OE-Prediction}(a), (c), and (d) illustrate the predicted yield stress, the  predicted Schmid factor for the operating slip system, and the slip mode, respectively. 
The inset in Fig. \ref{ch5:OE-Prediction}(a) quantifies the prediction accuracy by mapping the absolute error $|\sigma_{\rm y}^{\rm p}-\sigma_{\rm y}|$. 
Furthermore, Fig. \ref{ch5:OE-Prediction}(b) explicitly plots the predicted yield stress against the simulated values as a function of the loading polar angle, demonstrating the model's capability to capture the orientation dependent yield stress.

It can be inferred from a comparison with Schmid factor and activation mode in the original anisotropic activation polar diagrams (Fig. \ref{ch5:OE-computation-model}(c,d)) that soft mode deformation occupies a larger portion ($5^{\circ}<\theta<79^{\circ}$) of orientation space than expected purely on the basis of highest Schmid factor, and prevails in most of the orientation space originally attributed to hard mode I. This is a simple consequence of the fact that hard mode I has a much higher CRSS so that it is substituted with soft mode deformation. Indeed, according to the DDD results represented by the pole figure in Fig. \ref{ch5:OE-polar-figures}(c), soft mode deformation dominates in most of orientation space. At large polar angles, the area where hard mode II dominates plastic deformation also shrinks to a ring-shaped domain with loading polar angles $\theta>79^{\circ}$. 

\section{Conclusions and Outlook}\label{Conclusions-prospects}

In this paper, a DDD framework for twinned copper has been developed and applied to study the effects of loading orientation on dislocation microstructure evolution and mechanical behavior in columnar nanotwinned Cu, considering RVEs with a twin spacing $\lambda=160 ~{\rm nm}$ to investigate the evolution of dislocation-twin microstructures and study the resulting anisotropic deformation behavior under different loading orientations. The main results can be summarized as follows: 
If the loading axis is perpendicular to the stacked TBs, hard mode I dislocations activate and are confined between adjacent TBs, while drawn-out dislocations are deposited on both sides of the Tbs leading to strong hardening.  
If the loading axis is tilted to the stacked TBs, deformation is dominated by soft mode dislocations (TDs or lattice dislocations parallel to TBs), leading to a low yield stress and low hardening rate as activation of soft-mode dislocation sources is not impeded by the TBs. In this regime, twinning provides an alternative deformation mechanism that may contribute significantly to the overall deformation. 
If the loading axis is parallel to the TBs, yielding is affected by reactions between hard mode II dislocations and TBs so that a few hard mode II dislocations are transmitted. In this case, the TB decomposition of hard mode II dislocations leads to deposition of TDs.  

The interaction between dislocations and TDs can be described in terms of an effective CRSS that differs between the different deformation modes and can be used in a modified anisotropic Schmid law to predict the orientation dependence of slip system activation.
On this basis, an anisotropic model predicting for predicting slip system orientation in MTS has been formulated.

It is worth mentioning that we found that a high TD density with  an extremely important effect on the stress-strain response in MTSs
emerges when the loading axis is parallel to the TBs. Such a high density of defects at the twin boundaries, such as kink steps and Frank partial dislocations, has been observed in experiments \cite{Wang-Defective-twin-Nature-Mat-2013}. Therefore, the real-world defects of imperfect materials render the strengthening mechanisms more complex. In future work on size effects of nanotwinned materials, we will further explore the role of TB defects.

\section*{Acknowledgments}
This work was financially supported by the National Natural Science Foundation of China (Grant Nos. 12532004, 11672251, 12192214, 52192591),  Sichuan Science and Technology Program (Grant Nos. 2025HJRC0006, 2024NSFCJQ0068), and Basic Research Project of Taihang Laboratory (THL-K-24-115). MZ also acknowledges support by DFG under Grant No. Za 171 13/1.

\appendix
\renewcommand\thefigure{\Alph{section}\arabic{figure}}
\renewcommand\thetable{\Alph{section}\arabic{table}}
\setcounter{figure}{0}
\setcounter{table}{0}
\section{MD simulation of motion of individual dislocations in twinned copper}\label{MD-simulation}
\begin{figure}[htb]
	\centering
	\includegraphics[width=0.7\linewidth]{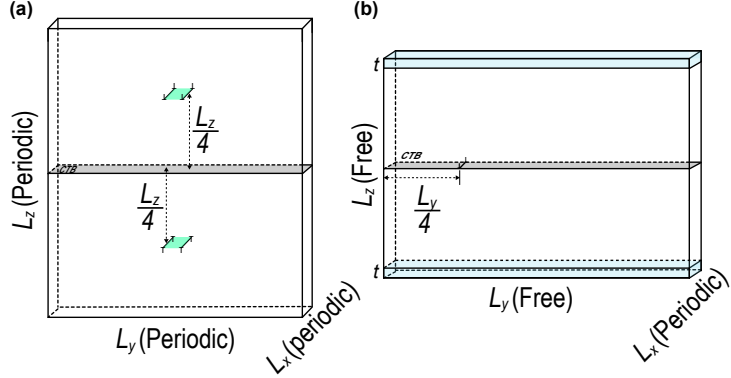}
	\caption{Computational models for MD simulations. (a) Full dislocation dipole, with stacking faults indicated in green; (b) Twinning dislocation, light blue denotes the loading (top) or constraint (bottom) layer atoms, with a thickness of $t$ = 5 nm.}
	\label{appendix:sketch-md}
\end{figure}

\begin{figure}[htb]
	\centering
	\includegraphics[width=0.8\linewidth]{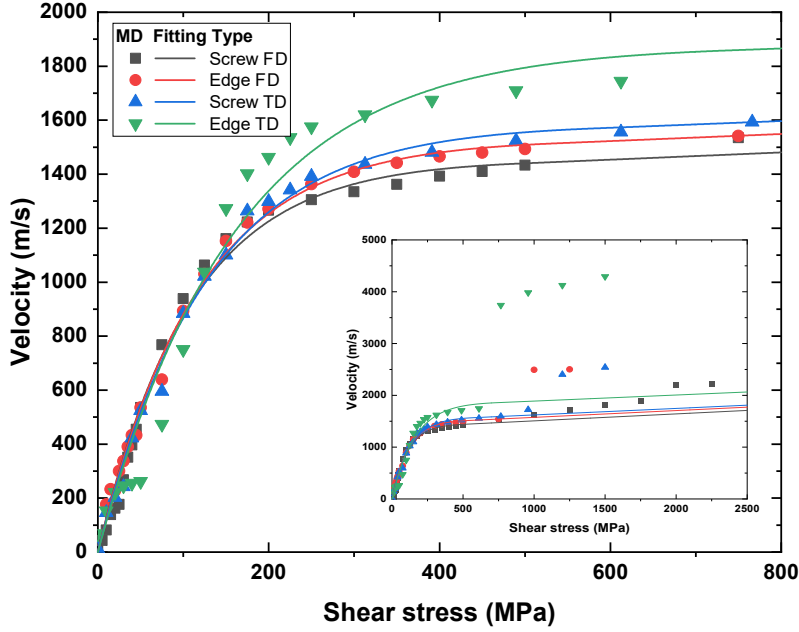}
	\caption{MD simulation results and fitting curves of velocity-shear stress relations for different types of dislocations, where FD denotes full dislocations and TD denotes twinning dislocations. The larger plot represents the relationship between velocity and stress at subsonic speeds, and the embedded plot represents the results for a wider range of shear stresses, which include transonic or hypersonic dislocation motion.}
	\label{ch2:velocity-shear-stress}
\end{figure}

\begin{table}[htb]
	\caption{Fitting parameters for the dislocation velocity-loading shear stress curves.}
	\label{ch2:table-fitting-parameter-mobility}	
	\centering
	\begin{tabular}{lll}
		\toprule
		Dislocation type & $v_{\rm m}~{\rm (m/s)}$ & $k~{\rm (10^{-9} ~Pa^{-1})} $  \\
		\midrule
		Screw FD & 1455 & 9.24 \\
		Edge FD  & 1529 & 8.58 \\
		Screw TD & 1586 & 7.91 \\
		Edge TD  & 1877 & 6.24 \\
		\bottomrule
	\end{tabular}
\end{table}
In this work, the kinematic behavior of individual dislocations is simulated by the open source MD platform LAMMPS \cite{Plimpton-introduction-lammps-JCP-1995}, using the embedded atomic method (EAM) potential established by Mishin et al. for copper \cite{Mishin-Structural-stability-Physical-2001}.

To establish the mobility laws for various types of dislocations in twinned copper, MD simulations were conducted on systems containing individual dislocations, specifically FDs and TDs. Four distinct dislocation types were examined: screw FD, edge FD, screw TD, and edge TD.

For the simulations involving full dislocations, a twinned supercell with  approximately 580,000 atoms is used as shown in Fig. \ref{appendix:sketch-md}(a), with the CTB normal consistent with Fig. \ref{ch5:OE-computation-model}(a). The dislocation line direction aligns with the $x$-axis, with a length $L_x$ = 3.1 nm. The glide direction follows the $y$-axis, with a length $L_y$ = 44 nm. The glide plane normal aligns with the $z$-axis, with a height $L_z$ = 50 nm. Periodic boundary conditions (PBC) are applied in all three directions.

The model incorporates a dislocation dipole consisting of two extended FDs with different glide planes. The distance between dislocations and CTB is set to $L_z/4$ to minimise the interaction force between the two dislocations.
Each dislocation is composed of two partials which are inserted into the supercell by imposing initially for each partial an approximate displacement field, which is obtained by integrating the density function of the Burgers vector in the non-singular dislocation model proposed by Cai et al. \cite{Cai-A-non-singular-Journal-2006}. 

In the geometric model for TDs, adjustments are made to the boundary conditions, box dimension, and dislocation positions, as illustrated in Fig. \ref{appendix:sketch-md}(b). PBC are applied exclusively along the line direction, while the boundaries in the other two directions represent surfaces. The length $L_z$ along the glide direction is increased to 256 nm, and the total number of atoms is raised to 2,900,000 to ensure a sufficiently long glide path before surface interactions dominate dislocation motion as the dislocation approaches the free surface. A single TDs is introduced at the CTB, situated at a distance of $L_y/4$ from the free surface $XOZ$.

All initial configurations are relaxed at 300 K for 50 ps in the NPT ensemble to obtain a stable structure. Subsequently, in-plane shear stresses are applied to the configurations for 25 ps´, with the shearing directions aligned with the dislocation Burgers vectors. For FD configurations that use fully PBCs, shear stresses are applied directly via the N$\sigma$T ensemble. For TD configurations, shear tractions and full constraints are applied, respectively, to the 5-nm-thick ($t$) atomic layers on the upper and lower surfaces, as shown in Fig. \ref{appendix:sketch-md}(b). To stabilize stress control and prevent stress oscillations, as initial condition a linear displacement profile (i.e. a pre-strain matched to the applied shear stress) is imposed on all atoms, starting from the relaxed structure.

In post-processing, common neighbor analysis (CNA) \cite{Tsuzuki_StructuralCharacterizationDeformed_2007} is used to identify the core region of the dislocations. 
In MD simulations at constant stress, the motion of both FD and TD dislocations  proceeded at approximately constant velocity, indicating that there is little influence of dislocation-dislocation interactions or free-surface image forces. Doubling the model size changed the dislocation velocities by less than 5\%.

Stress-velocity data are shown in Fig. \ref{ch2:velocity-shear-stress}. For FDs, the viscosity coefficient at the linear stage is approximately $15.6~\upmu{\rm Pa\cdot s}$, and the asymptotic terminal velocity is around 1,600 m/s. These values are consistent with those reported by Oren et al.\cite{Oren-Dislocationkinematicsmolecular-MaSiMSaE-2017}.
The study of nanotwinned copper materials in this work does not consider high strain rate loading, and the glide stresses in our DDD simulations hardly exceed $800~{\rm MPa}$, so only the mobility law in the range of subsonic velocities is developed. To obtain an analytical relationship between the dislocation velocity $v$ and the glide stress $\tau$ at subsonic velocities, an exponential equation describing velocity saturation can be fitted to the MD data \cite{Fan-Strain-rate-Nat-Commun-2021},
\begin{equation}
	v=v_{\rm m}\left(1-{\rm e}^{-k\tau}\right),
	\label{ch2:velocity-tau}
\end{equation}
where $v_{\rm m}$ and $k$ are fit parameters, which denote the saturation velocity and exponential slope, respectively. The fitting results for all four types of dislocations are shown in Table \ref{ch2:table-fitting-parameter-mobility}.

In the DDD simulations, dislocation velocities are evaluated by equating the nodal forces to opposing nodal drag forces which are obtained by inverting Eq. (\ref{ch2:velocity-tau}). Evaluation of the resulting nonlinear equations is done using a Newton iteration scheme. In this respect, Eq. (\ref{ch2:velocity-tau}) has the undesirable feature that it cannot be inverted when the velocity exceeds the limit value $v_{\rm m}$, hence the scheme cannot converge if a higher velocity appears in the course of the iterations. To mitigate this problem and ensure convergence, starting from a crossover velocity $r_{\rm c}v_{\rm m}$ where $r_{\rm c}=0.99$, the horizontal asymptote is replaced by a linear increase of the velocity with stress as shown in the inset of Fig. \ref{ch2:velocity-shear-stress}: 
\begin{equation}
	v=0.99 v_{\rm m} + 0.01 k v_{\rm m}(\tau - \tau_{\rm cr}), \qquad \tau > \tau_{\rm cr}
	\label{ch2:velocity-tau2}
\end{equation}
where the crossover stress $\tau_{\rm cr}$ obeys the relation $v(\tau_{\rm c})= 0.99v_{\rm m}$ as evaluated from Eq. (\ref{ch2:velocity-tau}). 

\section{Typical dislocation-TB interactions}\label{dislocation-TB-interaction}

\setcounter{figure}{0}
\setcounter{table}{0}

\begin{figure}[!htb]
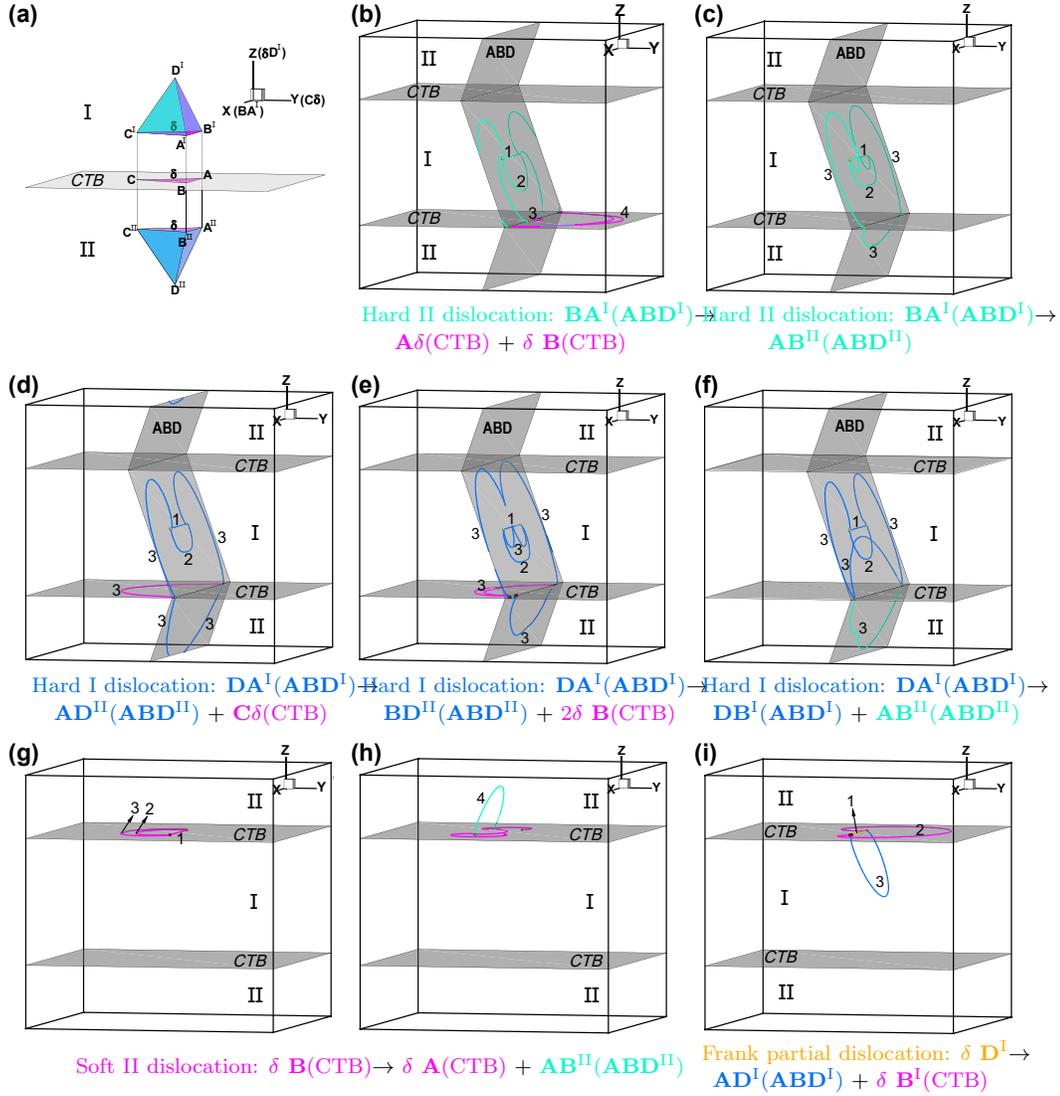

	\centering
	\begin{overpic}[width=\linewidth]{\linkDir/figures/chap2/sumary-interactions.pdf}

	\put(33,70.5){\scriptsize \textcolor[rgb]{0,0.9922,0.77647}{Hard II dislocation: $\textbf{BA}^{\rm I}(\textbf{ABD}^{\rm I})$}$\rightarrow$}
	\put(36,68){\scriptsize \textcolor[rgb]{0.9922,0,0.9922}{$\textbf{A$\delta$}({\rm CTB})$} $+$ \textcolor[rgb]{0.9922,0,0.9922}{$\textbf{$\delta$ B}({\rm CTB})$}}

	\put(64,70.5){\scriptsize \textcolor[rgb]{0,0.9922,0.77647}{Hard II dislocation: $\textbf{BA}^{\rm I}(\textbf{ABD}^{\rm I})$}$\rightarrow$}
	\put(70,68){\scriptsize \textcolor[rgb]{0,0.9922,0.77647}{$\textbf{AB}^{\rm II}(\textbf{ABD}^{\rm II})$}}

	\put(3,36.7){\scriptsize \textcolor[rgb]{0,0.4313725,0.9922}{Hard I dislocation: $\textbf{DA}^{\rm I}(\textbf{ABD}^{\rm I})$}$\rightarrow$}
	\put(5,34.2){\scriptsize \textcolor[rgb]{0,0.4313725,0.9922}{$\textbf{AD}^{\rm II}(\textbf{ABD}^{\rm II})$} $+$ \textcolor[rgb]{0.9922,0,0.9922}{$\textbf{C$\delta$}({\rm CTB})$}}

	\put(33,36.7){\scriptsize \textcolor[rgb]{0,0.4313725,0.9922}{Hard I dislocation: $\textbf{DA}^{\rm I}(\textbf{ABD}^{\rm I})$}$\rightarrow$}
	\put(35,34.2){\scriptsize \textcolor[rgb]{0,0.4313725,0.9922}{$\textbf{BD}^{\rm II}(\textbf{ABD}^{\rm II})$} $+$ \textcolor[rgb]{0.9922,0,0.9922}{$2\textbf{$\delta$ B}({\rm CTB})$}}

	\put(64,36.7){\scriptsize \textcolor[rgb]{0,0.4313725,0.9922}{Hard I dislocation: $\textbf{DA}^{\rm I}(\textbf{ABD}^{\rm I})$}$\rightarrow$}
	\put(65,34.2){\scriptsize \textcolor[rgb]{0,0.4313725,0.9922}{$\textbf{DB}^{\rm I}(\textbf{ABD}^{\rm I})$} $+$ \textcolor[rgb]{0,0.9922,0.77647}{$\textbf{AB}^{\rm II}(\textbf{ABD}^{\rm II})$}}

	\put(7,2){\scriptsize \textcolor[rgb]{0.9922,0,0.9922}{Soft II dislocation: $\textbf{$\delta$ B}({\rm CTB})$}$\rightarrow$  \textcolor[rgb]{0.9922,0,0.9922}{$\textbf{$\delta$ A}({\rm CTB})$} $+$ \textcolor[rgb]{0,0.9922,0.77647}{$\textbf{AB}^{\rm II}(\textbf{ABD}^{\rm II})$}}

	\put(64,3.2){\scriptsize \textcolor[rgb]{1,0.6667,0}{Frank partial dislocation: $\textbf{$\delta$ D}^{\rm I}$}$\rightarrow$}
	\put(65,0.7){\scriptsize \textcolor[rgb]{0,0.4313725,0.9922}{$\textbf{AD}^{\rm I}(\textbf{ABD}^{\rm I})$} $+$ \textcolor[rgb]{0.9922,0,0.9922}{$\textbf{$\delta$ B}^{\rm I}({\rm CTB})$}}

	\end{overpic}
	\caption{
	Interactions of single Frank-Read sources with TBs; probed by varying the loading axis such as to vary the shear stress on the outgoing planes. 
	(a) Schematic diagram of crystal orientations; 
	(b) Hard mode II dislocation decomposes into two soft mode II dislocations; 
	(c) Hard mode II dislocation transmits across the TB via cross-slip; 
	(d) Hard mode I dislocation  transmits across the TB and leaves a soft mode II dislocation; 
	(e) Hard mode I dislocation  transmits across the TB and leaves two soft mode II dislocations; 
	(f) Hard mode I dislocation  transmits across the TB by decomposing into a hard mode II dislocation moving into the twin and a hard mode I dislocation reflected back; 
	(g-h) Soft mode II dislocation bows out, emits a hard mode II dislocation into the adjacent grain and nucleates a new soft mode II dislocation; 
	(i) Frank partial dislocation decomposes into a hard mode I dislocation and a soft mode II dislocation.
	Line colors indicate different dislocation types as indicated below the figures. Numbers adjacent to dislocation lines indicate the evolution sequence.}
	\label{dislocation-TB-interactions}
\end{figure}

By introducing a single Frank-Read source into the simulation cell and varying the shear stress on the potential outgoing planes for dislocations transmitted through TBs, different types of dislocation-TB interactions and typical dislocation reaction products can be probed as shown in Fig. \ref{dislocation-TB-interactions}.

\bibliographystyle{elsarticle-num} 
\bibliography{\bibFile}





\end{document}